\documentclass[aps,rmp,reprint,amsmath,amssymb,graphicx,superscriptaddress]{revtex4-2}

\usepackage{bm}
\usepackage{xcolor}
\usepackage{hyperref}
\usepackage[normalem]{ulem}
\usepackage{graphicx}

\newcommand{\comment}[1]{}
\begin{document}

\title{\textit{Colloquium:} A critique on van der Waals and two-dimensional magnets}

\author{Johann Coraux}
\email{johann.coraux@neel.cnrs.fr}
\affiliation{Universit\'{e} Grenoble Alpes, CNRS, Institut NEEL, Grenoble INP, 38000 Grenoble, France\looseness=-3}

\author{Nicolas Rougemaille}
\affiliation{Universit\'{e} Grenoble Alpes, CNRS, Institut NEEL, Grenoble INP, 38000 Grenoble, France\looseness=-3}

\author{Cedric Robert}
\affiliation{Universit\'{e} de Toulouse, INSA-CNRS-UPS, LPCNO, 135 Av. Rangueil, 31077 Toulouse, France\looseness=-4}

\author{Cl\'{e}ment Faugeras}
\affiliation
{Laboratoire National des Champs Magn\'{e}tiques Intenses, LNCMI-EMFL, CNRS UPR3228, Univ. Grenoble Alpes, Univ. Toulouse, Univ. Toulouse 3, INSA-T, Grenoble and Toulouse, France\looseness=-7}

\author{Andr\'{e}s Saul}
\affiliation{Aix-Marseille Universit\'{e}, CNRS, CINaM, 13288 Marseille, France\looseness=-4}

\author{Beno\^{i}t Gr\'{e}maud}
\affiliation{Aix Marseille Universit\'{e}, Université de Toulon, CNRS, CPT, Marseille, France\looseness=-5}

\author{Luis Hueso}
\affiliation{CIC nanoGUNE BRTA, Donostia–San Sebastian 20018, Spain\looseness=-4}
\affiliation{IKERBASQUE, Basque Foundation for Science, Bilbao 48009, Spain\looseness=-4}

\author{F\'{e}lix Casanova}
\affiliation{CIC nanoGUNE BRTA, Donostia–San Sebastian 20018, Spain\looseness=-4}
\affiliation{IKERBASQUE, Basque Foundation for Science, Bilbao 48009, Spain\looseness=-4}

\author{Aur\'{e}lien Manchon}
\email{aurelien.manchon@univ-amu.fr}
\affiliation{Aix-Marseille Universit\'{e}, CNRS, CINaM, 13288 Marseille, France\looseness=-4}

\date{\today{}}

\begin{abstract}

Magnetic two-dimensional (2D) crystals were isolated about a decade ago, triggering a tremendous research activity worldwide. This colloquium raises a stiff question: what is really new about them? At first sight, they seem to be purer implementations of 2D spin models than traditional systems such as ultra-thin films. Yet, they partly realized their promises so far, and whether they give fresh perspectives on long-standing predictions in statistical physics is still an open question. Undoubtedly, they are uniquely amenable to electric-field effect, susceptible to mechanical deformation, and sensitive to moir\'{e}s. They are interesting platforms for exploring, challenging, or simply revisiting a wide range of phenomena in condensed matter magnetism. This colloquium proposes a critical, yet not necessarily skeptical, overview of the field, clarifying what we believe could be unique with 2D magnets, related quasi-2D van der Waals magnets, and their heterostructures.

\end{abstract}


\maketitle

\tableofcontents{}

\section{Preamble}
\label{sec:preamble}

\begin{figure*}[!hbt]
\includegraphics[width=157.76mm]{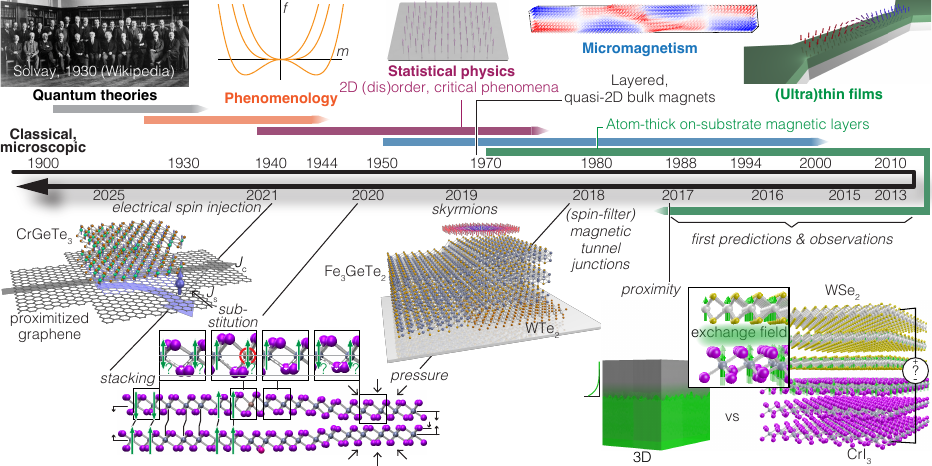}
\caption{\label{fig:timeline}Top part: Historical approaches to solid-state magnetism. Arrows tentatively delimit the early development of each field. Bottom part: Illustration of some milestones in the study of 2D/vdW magnets. The role of relative atomic stacking between successive layers (altered by planar strains, lateral sliding, or moir\'{e}s), of atomic substitution or applied pressure, on the orientation/length of interatomic bonds, and consequences on the superexchange interaction between spins, is highlighted at the bottom-left. Center: a heterostructure with a high-spin-orbit-interaction 2D material (WTe$_2$), and the possible thus-generated skyrmion in Fe$_3$GeTe$_2$ \cite{Yang2020b}. Right: magnetic proximity effect, with some spin polarization induced in the top semiconductor and a weakening of magnetism is the ferromagnet underneath (the situation for non-vdW 3D materials illustrates issues with a rough interface creating a marginal interfacial effect). Proximity is either at equilibrium (absence of external driving forces, e.g., voltage source or laser beam) or not. Left: spin-polarized current $J_\mathrm{s}$ through graphene proximitized by a vdW magnet [CrSBr \cite{Ghiasi2021}, CrGeTe$_3$ \cite{Yang2025}] and flowed by a charge current $J_\mathrm{c}$.}
\end{figure*}

Two-dimensional (2D) crystals became widespread in the mid-2000s, when a since-then popular mechanical exfoliation process, combined with a straightforward optical identification, gave access to their unique physical properties. Magnetic 2D crystals were isolated a decade later, in 2016-2017, at a time when solid-state magnetism had long gone beyond low-dimensional phenomena, actually exploiting them towards advanced functionalities. Yet, multiple key questions of 2D magnetism remain unanswered today, related for instance to the nature of the spin interactions, predictions from statistical physics models, or continuous descriptions of low-dimensional magnetism. Moreover, revisiting 2D magnetism through the prism of strong correlation effects, topological concepts, or collective quantum phenomena has proven to be very rich in the past two decades. Still, one may legitimately wonder what 2D magnets specifically bring in all these, beyond merely offering a new class of materials.

Some decades ago, ultra-thin and multi-layered magnetic materials sparked off the development of electronic devices exploiting the spin degree of freedom, and the field of spintronics more specifically. Similarly, today's 2D magnets and their heterostructures are being considered for a new generation of devices. Whether these will perform better than, or differently from, architectures made of more traditional materials, whose figures-of-merit have been pushed very high at room temperature, needs to be discussed.

By now, the tone of our review is probably clear: in a context where 2D magnets have turned into a mainstream research topic, we will be playing devil’s advocate. We think it is insightful to challenge the common perception that 2D magnets are disruptive playgrounds to revisit and renew 2D magnetism. For such a fast-moving field of research, which has seen its share of dubious claims and debates, sometimes progressing with gaps of knowledge on fundamental magnetism concepts, we hope that our point of view is helpful. Our aim in taking this (inevitably background-dependent) perspective on the scientific literature is to help identify what is truly unique about 2D magnets and related layered compounds, and to clarify what developments may still lie ahead. This Colloquium is primarily devoted to students and physicists working with 2D materials without a background in magnetism; it should also be informative for magnetism experts (who might skip Sec.~\ref{sec:lessons} and directly jump to Sec.~\ref{sec:whatsnew}) to figure out what has been and could be studied with these (somehow) new materials.

Multiple review articles and viewpoints have been issued already, including early on \cite{Burch2018,Gibertini2019}. Here we do not aim at comprehensiveness, and refer the readers seeking further information to more extended reviews, e.g., \cite{Wang2022,Park2025}. In Sec.~\ref{sec:lessons} we will recall (a necessarily biased selection of) what was already known before 2016–2017. Some of this material has been discussed, at least in part, in earlier reviews and in textbooks. Our goal is to offer additional perspectives. In Sec.~\ref{sec:whatsnew}, we will highlight new physical effects. Since such effects rarely emerged entirely serendipitously, we will, whenever relevant, recall in what sense they had been anticipated or partially explored in other materials. We will also address several active questions in magnetism---such as frustration physics, orbital magnetism, topology and quantum geometry---whose exploration with 2D magnets has so far remained limited.

The meaning of `2D' is often ambiguous in the literature, as noted early on \cite{Mills1991}. A system behaves effectively two-dimensionally when two spatial directions suffice to describe its physical properties. This typically occurs when the characteristic length scale $\lambda$ of the phenomenon (e.g., mean free path or magnetic exchange length) largely exceeds the thickness. For van der Waals (vdW) materials, `2D' usually refers to a single stable atomic layer, although electrons and spin densities still possess a finite extension out of plane. Consequently, properties can vary strongly between one and, say, 10 monolayers, a thickness range often associated with 2D materials \cite{Geim2007}. In magnetism, the term is commonly linked to a universal behavior around a critical point expected in atomically thin layers, though truly 2D magnetic behavior is rarely realized experimentally. Moreover, even atomically thin magnets remain embedded in a three-dimensional environment. Substrates, capping layers, heterostructures, and ripples \cite{Meyer2007,Geringer2009} all prompt the possibility of achieving a strict 2D limit. Nevertheless, one- and few-layer magnets have recently revived interest in 2D magnetism and renewed attention toward thicker vdW magnets, often termed `quasi-2D'. In these materials, individual layers can behave nearly independently and support magnetic excitations with negligible dispersion along the stacking direction. This colloquium addresses both atomically thin and thicker vdW magnets (grouped here as `2D/vdW'), whose studies are now closely connected. While atomically thin magnets still present major experimental challenges, thicker vdW systems likely account for most work published since 2017.

\section{Foundations and prior knowledge on 2D magnetism}
\label{sec:lessons}


This section is intended as a `survival kit' for newcomers, skimming the century-long development of condensed-matter magnetism (see Fig.~\ref{fig:timeline}) and raising points often overlooked in the 2D/vdW-magnetism literature: What is specific about spin interactions in vdW systems? What is distinctive about magnetic order and disorder in 2D? Why do some ferromagnets show no macroscopic magnetization, and how does this relate to domain walls? What were the earliest experimental signatures of 2D magnetism? How do strong electronic correlations give rise to ordered quantum phases? What are the hallmarks of topology and quantum geometry? 

\subsection{Microscopic quantum spin interactions}
\label{ssec:micro}

Solid state magnetism is fundamentally quantum: the Bohr–van Leeuwen theorem states that classical statistical physics cannot account for either spontaneous magnetization or field-induced paramagnetism \cite{Fazekas1999}. Microscopically, magnetic moments arise from the electron’s spin and orbital angular momentum. In crystals, these moments reside on ions subjected to the surrounding charge distribution---whose crystal field lifts orbital degeneracies and produces single-ion (magnetocrystalline) anisotropy when a spin-orbit interaction is present---as well as to the surrounding distribution of magnetic moments, through the dipole–dipole interaction whose strength is typically $\frac{\mu_0 (g \mu_B S)^2}{4\pi r^3} \approx 0.16$~K = 14~$\mu$eV, for $S=1$ spins at a distance $r = 2.5$~\AA. This small interaction, although always present, obviously does not account for the diversity of magnetic orders observed at most experimentally accessible temperatures. The prevalent spin interactions in solids are of different microscopic origins, all quantum in nature. 

\begin{figure}[!bt]
\begin{center}
\includegraphics[width=80mm]{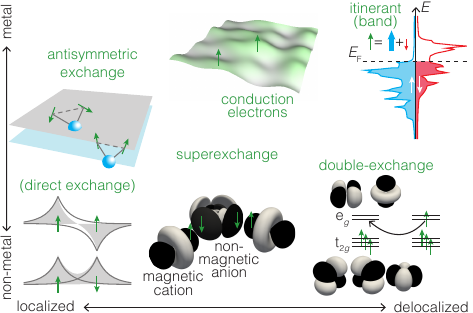}
\caption{\label{fig:interactions}Spin-spin interactions, in the picture of localized spins and in the paradigm of itinerant magnetism. Direct exchange (rarely relevant in practice), anisotropic DMI, superexchange (in metals, via conduction electrons, by Ruderman-Kittel-Kasuya-Yoshida interactions, or in semiconductors/insulators, via non-magnetic ions), and double-exchange (multi-valence compounds) are represented.}
\end{center}
\end{figure}

The first contribution is direct exchange, originating from Coulomb electron–electron repulsion and the Pauli principle enforcing antisymmetry of the electronic wave function. Its sign depends on the overlap between the one-electron orbitals carrying the magnetic moments. When this overlap is small---as for large interatomic distances or orthogonal orbitals on the same ion---direct exchange favors ferromagnetic alignment; relaxing orthogonality may instead yield antiferromagnetic coupling. Other exchange-type mechanisms, such as kinetic and double exchange (Fig.~\ref{fig:interactions}), are often stronger. In many 2D/vdW magnets, superexchange via non-magnetic ions \cite{Zhang2019} plays a central role; its trends are captured by the Goodenough–Kanamori–Anderson rules \cite{Goodenough1963}, which relate bond angles and orbital character to the sign and magnitude of the interaction. Higher-order terms, e.g., biquadratic and four-spin exchange, may also matter at low temperatures \cite{Takahashi1977,Kartsev2020} and promote noncollinear magnetic order. Overall, these mechanisms typically favor antiparallel spin alignment, consistent with the prevalence of antiferromagnets in nature. Breaking of inversion symmetry together with spin-orbit coupling, in bulk or at interfaces, can produce antisymmetric exchange, named Dzyaloshinskii-Moriya interaction (DMI), conducive to a variety of noncollinear spin alignments \cite{Dzyaloshinskii1957,Moriya1960}, including, e.g., single-$q$ and multiple-$q$ spin spirals.

Modeling magnetic interactions from first principles is challenging, as no single electronic-structure method can yet treat electronic correlations in a universally reliable way \cite{Szilva2023}. Standard density functional theory (DFT) within the local spin-density (LSDA) or generalized gradient (GGA) approximations is often insufficient for transition-metal compounds, prompting the widespread use of a Hubbard $U$ correction 
\cite{Pakdel2025}. This parameter depends strongly on the chemical environment and oxidation state. It is frequently chosen empirically, and cannot capture non-local correlations, such as found in itinerant magnets. Hybrid functionals, e.g., Heyd–Scuseria–Ernzerhof (HSE06), partly remedy these limitations by mixing LSDA/GGA with Hartree–Fock exchange, yielding more accurate band structures for 2D/vdW magnets, albeit at a higher computational cost \cite{Wang2022}. Once electronic structures are computed, exchange interactions can be extracted and are often interpreted using simplified ligand-field or tight-binding–based models \cite{Khomskii2014}, which incorporate orbital energies, crystal-field effects, hopping integrals, and Hubbard-like Coulomb and exchange terms. 
Noteworthy, the effective exchange interaction does not necessarily weaken monotonically with distance. For example, in MnPS$_3$, the exchange interaction between second nearest-neighbors almost vanish \cite{Wildes1998,Sivadas2015,Vaclavkova2020}, whereas in CrOCl it survives up to the seventh nearest-neighbors \cite{Pawbake2025}. Finally, note that since vdW interactions are not naturally captured in standard DFT, they must be taken into account through explicit corrections for layered magnets.

The structural anisotropy of vdW magnets poses a fundamental question about the strength of interlayer spin interactions. An exchange-like scenario (e.g., super-superexchange) generally assumes mediation by at least two non-magnetic ions and electron hopping through the vdW gap --- e.g., the Cr-I-vdW-I-Cr path in CrI$_3$. Such an interlayer exchange interaction is several orders of magnitude smaller than the leading intra-layer exchange interactions. In CrSBr, which exhibits a N\'{e}el temperature of $\sim$130~K \cite{Lee2021}, intra-layer (ferromagnetic) interactions are 20-80~K = 2-7~meV \cite{Wang2023}, while the interlayer (antiferromagnetic) interaction is about 0.07~K = 0.006~meV \cite{Cho2023}. Actually, the latter interaction, being a higher-order perturbation-theory term in essence \cite{Blundell2001}, is very often too small to be evaluated reliably with the level of precision offered by DFT methods. The very small value is however not antinomic with interlayer magnetic ordering, as the extended 2D order (within each layer) would otherwise imply cumulatively strong penalty for interlayer orders not complying with the sign of the interlayer interaction. In practice, despite interlayer interactions in the (sub-)Kelvin range, order in the perpendicular direction persists at much higher temperatures. However, compared to more traditional (3D covalent) systems, the spin dynamics is expectedly much softer in this direction.

Many physical phenomena addressed experimentally are classical ones (see next two subsections) occurring well beyond the atomic scale. It is then desirable to shift from the localized picture of an on-lattice quantum spin Hamiltonian\comment{, e.g., of the form $-\sum_{\langle i,j \rangle} J\;\hat{S}_i \cdot \hat{S}_j$ (with $\hat{S}_{i,j}$ spin operators for first neighbours on the lattice sites $i,j$, at a distance $a$ and coupled by an exchange interaction $J$)}, to a continuous description, based on a vector field of classical magnetic moments ${\bf m}$ (normalized by the saturation magnetization, $M_s$). The field's energy, being proportional to $\frac{2JS^2}{a} \int [(\bm{\nabla} m_x)^2+(\bm{\nabla} m_y)^2+(\bm{\nabla} m_z)^2]d^3r$, with $a$ the inter-spin distance and $J$ the exchange interaction, accounts, at the same time, for the stabilization of the ground-state magnetic order, and for the energetic cost of deviations from a uniform magnetization \cite{Hubert1998}.

The relevance of a description in terms of localized spins, when dealing with conduction electrons in crystals where they propagate (delocalized electrons) as is the case of metallic materials, is questionable. Alternative descriptions, e.g., Stoner's model of itinerant magnetism (Fig.~\ref{fig:interactions}) may help rationalize the observations, although their predictive power is often limited. Alternatively, the free energy variation associated with small magnetic fluctuations in metals can be expressed as a sum of two terms, one effectively corresponding to the creation of on-site magnetic moments and the other to the interaction between these moments \cite{deLacheisserie2005}. This is strikingly reminiscent of the localized-spin picture discussed previously.

\subsection{Statistical physics of 2D magnetism}
\label{ssec:stat}


Statistical physics has been a workhorse for over a century to study order and disorder in magnetism, and magnetism in turn has stimulated the development of many theoretical approaches and methods. This reciprocity arises partly from the diversity of spin system (Ising, Potts, XY, Heisenberg, classical or quantum) and the wide range of relevant geometries and dimensionalities. By the 1950s, a key question was whether statistical physics could treat models capable of explaining magnetic ordering \cite{Onsager1944}. This view translated in a far-reaching concept, that magnetic order emerges as a collective phenomenon in interacting spin systems. As Fisher later emphasized, 2D models, crucial for understanding critical behavior, can be ``well-approximated physically by carefully chosen \textit{layered magnetic systems}'' \cite{Fisher1974}—in modern terms, vdW magnets!

Beyond Onsager's exact solution on Ising spins, which indeed predicts order, lattice geometry plays a crucial role. The triangular lattice, with Ising spins in antiferromagnetic interaction, provides a seminal example of a system that does not order and remains liquid-like even at zero temperature, for both classical \cite{Wannier1950,Houtappel1950} and quantum \cite{Fazekas1974} spins. But a specific geometry is not the only ingredient for disorder. With XY spins in ferromagnetic interactions, no long-range order can exist either, and the system is stabilized by topological defects \cite{Kosterlitz1973}. If the spins are now of Heisenberg type and in first-neighbor interaction, Mermin-Wagner fluctuations destroy long-range order too \cite{Mermin1966}. In fact, even in 3D (in the thermodynamic limit and without applied field), due to dipole-dipole interactions, spontaneous magnetization is predicted to vanish exactly \cite{Griffiths1968}. This remarkable absence of spontaneous magnetization mirrors the lack of long-range order predicted for isotropic 2D planar and 2D Heisenberg ferromagnets. Looking in the rear-view mirror, one may provocatively reformulate the initial idea: is disorder the general trend in magnetism rather than order, in 2D?

\begin{figure}[!bt]
\begin{center}
\includegraphics[width=80mm]{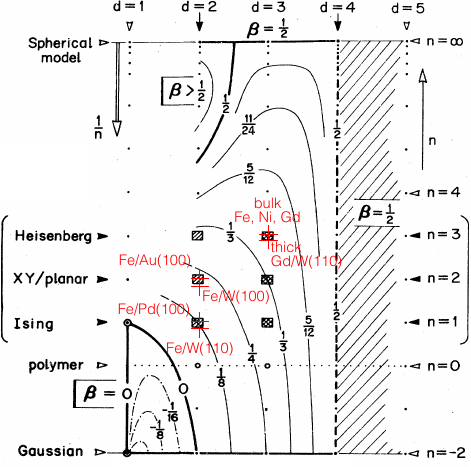}
\caption{\label{fig:phystat}Contour of constant magnetization exponent $\beta$ in the $(d,n)$ plane ($d$: dimensionality; $n$: spin degree of freedom). From~\cite{Fisher1974}. The $\beta<0$ region is nonphysical. In red: experimental data for more-or-less thin films / bulks \cite{Vaz2008}.
}
\end{center}
\end{figure}

In Gibbs’ approach, one considers multiple copies of a system composed of a large collection of spins to derive averages and variances of the system's energy ($E$) and magnetization ($M$). From these, experimental observables such as the susceptibility $\chi$ and specific heat $C$ are obtained as a function of temperature $T$. Across phase transitions, standard deviations may become non-negligible, and even diverge, signaling fluctuations of the order parameter at all length scales. Such critical phenomena are described by classes of Hamiltonians exhibiting universality, where critical exponents and scaling behavior are independent of microscopic details \cite{Fisher1974}. This idea has been central in the development of theories \cite{Domb1949,Kadanoff1966,Fisher1967,Wilson1971}, especially based on the renormalization group, that account for experimental critical exponents of, e.g., $M(T)$ (Fig.~\ref{fig:phystat}), $\chi(T)$, or $C(T)$.


While early measurements focused on the evolution of the Curie temperature, $T_{\mathrm{Curie}}$ \cite{Suter1979}, and $M_\mathrm{s}$, as the monolayer regime is approached \cite{Gradmann1968,Neugebauer1959}, the question of 2D universality classes was attacked experimentally in the late 1980s. This became feasible only once surface science techniques matured, and atomically clean, crystalline magnetic monolayers could be grown and probed with high sensitivity \cite{Gradmann1993}. In retrospect, experiments back then were demanding and subject to debates \cite{Cortie2020}, with sometimes unexpected or even conflicting results \cite{Rau1989,Liu1990,Qiu1994,Benvcok2006}, related to the way some measurements were performed, different possible kinds of growth or to the non truly-2D nature of the films (thicknesses from 1 to several 10 layers were studied). Still, some of these early experiments remarkably demonstrated that certain transition metal or rare-earth films exhibit genuinely 2D critical exponents \cite{Farle1987,Liu1990,Qiu1994,Kohlhepp1992,Back1995,Elmers1996}. At this time, determining the spin symmetry (Ising, Heisenberg, XY) and understanding substrate effects, including interface magnetic anisotropies, were central challenges.

In favourable cases, data showed clear trends. The $\beta$ critical exponent---in $M\propto(1-T/T_\mathrm{Curie})^\beta$---was found to approach or even reach three specific values corresponding to well-known universality classes (Fig.~\ref{fig:phystat}): 2D Ising, e.g., in Fe on Pd(100); 2D XY, in Fe on Au(100); 3D Heisenberg in thick films and bulks \cite{Vaz2008}.
Noteworthy, a few studies back then addressed the thickness dependence of the critical exponents, highlighting a 3D/2D dimensional crossover \cite{Li1992,Huang1993,Huang1994}.

Experiments do not always conclusively elucidate the universality class, as measured critical exponents can scatter around or deviate from theoretical predictions. For instance, using the temperature dependence of remanent magnetization as the order parameter can lead to inaccurate estimates of $T_\mathrm{Curie}$ and $\beta$, especially in systems with magnetic domains. It may instead be advantageous to track magnetization under a small applied field \cite{Poulopoulos1997}. Moreover, the actual spin symmetry is never strictly speaking, e.g., Ising or XY, but 3D in nature with varying degrees of anisotropy, which can be probed experimentally through magnon dispersion \cite{Zhu2025}. Besides, the thickness of ultrathin films on substrates is rarely uniform, and intermixing can occur at interfaces. Substrate effects---electronic hybridization, heteroepitaxial stress, magnetic polarization, work function differences, and spin reorientation---can all modify the magnetic moment and anisotropy, making each layer/substrate system effectively unique. 

Are 2D/vdW magnets immune to interface-related issues? To some extent, yes, yet we will see in Sec.~\ref{ssec:2Dmag} that they still do not allow to achieve a fully settled picture on 2D universality classes. 

\subsection{Micro- and nano-magnetism}
\label{ssec:micronano}

Whether magnetic order always persists or may vanish under certain conditions is a key question for bulk magnets, just as it is for the 2D on-lattice spin models discussed in the previous section, although from a different perspective \cite{Kittel1949}. It was in fact not clear at the beginning of the 20th century why a ferromagnetic body could carry no macroscopic magnetization at $T<T_\mathrm{Curie}$, or why magnetization could be entirely washed out in presence of a very small magnetic field. Of course, we are nowadays familiar with the concepts of internal molecular field and domain walls \cite{Weiss1907}, confirmed theoretically \cite{Heisenberg1928,Landau1935} and experimentally as the first magnetic imaging techniques were invented \cite{Bitter1931,Hamos1931,McKeehan1934}. Since then, directly seeing magnetic domain and domain walls has become possible in multiple ways, with exquisit refinement in terms of spatial (atomic-scale) or temporal (sub-ps) resolution, sometimes under extreme conditions of temperature, pressure and magnetic field. Technical developments are so advanced that the nanomagnetism and spintronic communities turn their attention to 3D imaging for the reconstruction of complex magnetization textures embedded within a volume \cite{Donnelly2017}. Research on 2D/vdW magnets has only exploited magnetic imaging to a limited extent so far [still, providing experimental demonstration of 2D ferromagnetism \cite{Gong2017,Huang2017}].

In most cases, magnetic imaging correlates particularly well with micromagnetic predictions, i.e., a continuous description of the magnetization vector field whose dynamics obeys an equation of motion. 
In its simplest form, this equation takes into account the precession of the local magnetization about the total effective field (external plus demagnetization fields), and a damping term that reflects energy dissipation and the eventual alignment of magnetization along the effective field. The 2D limit, especially when the layer is decoupled from its environment (substrate, capping, etc), raises an interesting question: is it fully equivalent to model their properties using on-lattice (i.e., discretized) spin Hamiltonians or with the (continuous) Landau-Lifshitz-Gilbert equation of micromagnetism? Efforts to relate this equation to a microscopic viewpoint build on \textit{ab initio} calculations accounting for different kinds of spin interactions and sources of magnetic anisotropy \cite{Wahab2021a,Wahab2021b,Alliati2022,Yang2022}.

Provided that the magnetization texture is at thermal equilibrium [this is not always for granted, including for certain vdW compounds \cite{Wahab2021a}], the nature of a domain wall (N\'{e}el, Bloch or other) and its internal structure reflect the competition between (anisotropic) exchange, demagnetization and/or magnetocrystalline anisotropy energies. For instance, the square of the domain walls' width is related to the ratio of the exchange energy to the anisotropy energy (which is generally ill-defined), and ranges typically between 10 and 100~nm for common metals and alloys. So far, it remains unclear whether the studied static domain walls, and the magnetic interactions at play, qualitatively differ from those encountered in non-vdW materials \cite{Purbawati2020,Yang2022,Zur2023}.  

The dynamics of the domain walls, induced by an external field or a spin current, also seems globally reminiscent of what is known in other materials \cite{Beach2008,Manchon2019}, pointing to a similar role of the different mechanisms behind domain wall motion (spin-transfer or spin-orbit torques) \cite{Zhang2024a}. There might be distinctive, nonlinear dynamical effects, yet their occurrence seems coincidental, for certain vdW compounds with specific combination of spin damping, absence of conduction electrons, and weak magnon scattering \cite{Wahab2021b}. The role of atomically well-defined edges on the domain wall dynamics has been pointed out \cite{Wahab2021b,Alliati2022}, but controlling their crystallography does not seem more straightforward in 2D/vdW ribbons than in more standard materials. Other crystal imperfections influence, e.g., magnetization reversal, and their role is usually studied with magnetometry. The latter, however, lacks sensitivity for small magnetic objects such as an exfoliated flake, in which case imaging is an appealing alternative, revealing which of the two standard mechanisms is relevant, nucleation or pinning of domain walls \cite{Broadway2020}.


Finally, the absence of spontaneous magnetization predicted in infinite 3D systems with dipole-dipole interactions \cite{Griffiths1968} also applies to 2D systems \cite{Pescia2024}. Studying a series of exfoliated samples with controlled thickness and lateral dimensions could reveal the mechanism by which magnetic domains emerge in 2D/vdW ferromagnets.

\subsection{Early signatures of 2D spin excitations}
\label{ssec:historicalmagnon}

The arguments ruling out long-range order in 2D lattices of Heisenberg spins with short-range interactions (see Sec.~\ref{ssec:stat}) raised questions about the central role of spin excitations \cite{Stanley1966}. Their investigation was accelerated in the late 1960s by the advent of neutron scattering and Raman spectroscopy, and later by (resonant) inelastic X-ray scattering and additional techniques (see Sec.~\ref{ssec:topoMagnons}). Using these tools, experimentalists soon discovered that certain layered 3D compounds display magnetism with strong 2D character.

\begin{figure}[h!]
\includegraphics[width=8cm]{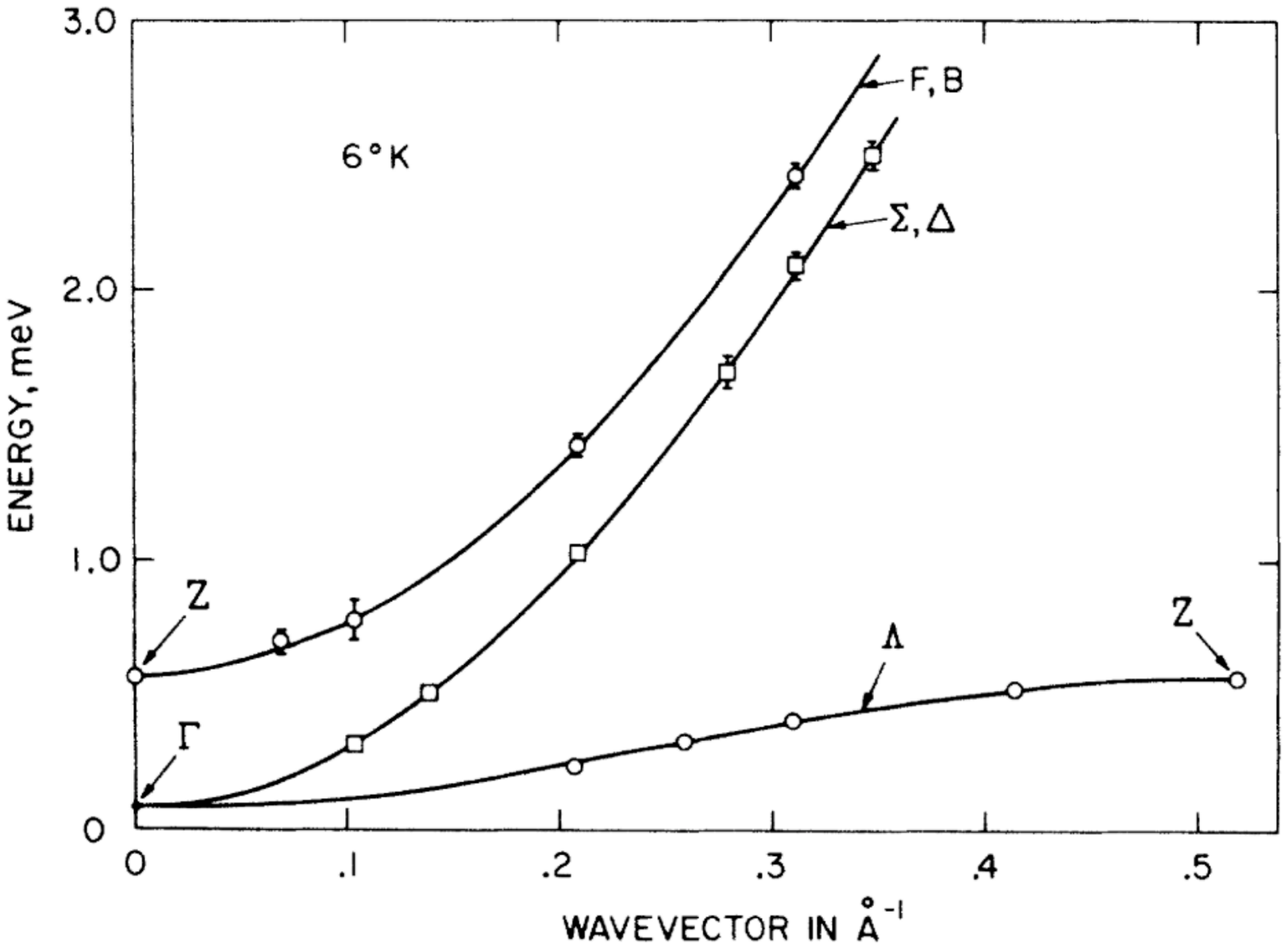}
\caption{Magnon dispersion of CrBr$_3$ measured by inelastic neutron scattering along the $\Gamma$-$\Delta$, Z-F (in-plane) and $\Gamma$-Z (out-of-plane) directions. The strong anisotropy in the magnon dispersion advocates for 2D magnetism. From \cite{Samuelsen1971} \label{fig:Sam71}}
\end{figure}

CrBr$_3$ received particular attention in this context \cite{Tsubokawa1960,Davis1964}. Inelastic neutron scattering revealed strongly anisotropic magnon dispersion and renormalization, such that bulk CrBr$_3$ is since then considered a quasi-2D Heisenberg ferromagnet \cite{Samuelsen1971} (see Fig.~\ref{fig:Sam71}). At about the same time, a Ruddlesden–Popper compound, K$_2$NiF$_4$, was proposed as a quasi-2D Heisenberg antiferromagnet, based on neutron scattering data featuring rod-like, elongated Bragg ``peaks'' \cite{Birgeneau1969}, a flat magnon dispersion along the $c$-axis \cite{Skalyo1969}, and the temperature dependence of the energy of magnon pairs \cite{Fleury1970}. The latter was also addressed in several other compounds of the same family, like Rb$_2$(Fe,Mn)F$_4$ \cite{Birgeneau1970a,Birgeneau1973,Birgeneau1977} and RbNiF$_3$ \cite{AlsNielsen1972}.

In the mid-80s, the discovery of high-temperature superconductivity in layered cuprates prompted interest in their magnetic order \cite{Anderson1987}. Their lamellar structure, consisting of stacked CuO$_2$ planes, advocates for 2D magnetism. Starting with the observation of finite-range antiferromagnetic spin correlations in the paramagnetic phase of La$_2$CuO$_4$ \cite{Shirane1987}, neutron and Raman scattering experiments then confirmed that the CuO$_2$ spin planes behave as 2D Heisenberg antiferromagnets with $S$=1/2 \cite{Kastner1998}.

Before the emergence of 2D magnets (i.e., single layers) and beyond Heisenberg systems, spin excitations have also been scrutinized in quantum spin liquid candidates \cite{Mendels2016}, including herbertsmithite and $\alpha$-RuCl$_3$. The former is expectedly a practical implementation of the antiferromagnetic kagome model \cite{Norman2016}. Neutron scattering experiments unveiled a continuous spectrum of spin excitations, compatible with the quantum spin liquid scenario \cite{Han2012,Punk2014}, a kind of signature that was also sought for in similar compounds \cite{Bonbien2022}. $\alpha$-RuCl$_3$ is a 3D vdW material with strong bond-dependent exchange, often considered a proxy for a Kitaev antiferromagnet \cite{Plumb2014}. Inelastic neutron scattering has revealed possible signatures of the Majorana excitations predicted for such systems \cite{Banerjee2017}, although the debate remains largely open.

\subsection{Strong electronic correlation effects}
\label{ssec:correl}

In condensed matter, the Coulomb interaction between electrons, primarily through Pauli's exclusion principle, generally has a weak effect on transport properties. This is well accounted for by Fermi liquid theory, particularly at intermediate temperatures where quantum coherence develops, albeit only over short length scales~\cite{Landau1957,Marder2010}; a classic example is $^3$He~\cite{Leggett2004}. However, this Fermi liquid phase is usually unstable under small perturbations (electronic density, external pressure, stoichiometry, or simply by lowering temperature), inducing new quantum phases with radically different properties. For instance, triplet $p$-wave superconductivity (similar to $^3$He) emerges in Sr$_2$RuO$_4$ out of a Fermi liquid upon decreasing the temperature~\cite{Stricker2014}, and the La$_{2-x}$Sr$_x$CuO$_4$ cuprate undergoes a transition from an antiferromagnetic state to a high-$T_\mathrm{c}$ superconducting phase upon hole doping~\cite{Scalapino2012,Keimer2015}. The instability of a regular metallic phase towards complex, possibly competing phases is the hallmark of strongly correlated materials~\cite{Morosan2012,Georges2024}.

It turns out that most of these strongly correlated phenomena appear in materials with strong 2D character, for several reasons: (\textit{i}) confining electrons at reduced dimension enhances the relative strength of the Coulomb interaction compared to the kinetic energy; (\textit{ii}) the impact of temperature and/or quantum fluctuations becomes more prominent when lowering dimensionality from 3D to 2D (for a given kind of spin degree of freedom, e.g., Heisenberg, XY, Ising), making phases proner to instability; (\textit{iii}) topological effects naturally emerge in 2D. 

How thermal/quantum fluctuations impact the complexity of a phase diagram finds a striking illustration in cuprates, which feature many competing phases ($d$-wave superconductivity, spin and charge density waves, Fermi liquids, etc), resulting in fluctuations or precursor orders with marked signatures even at temperatures above the transitions~\cite{Seibold2021}. More recently, experiments involving vdW transition metal dichalcogenides have unveiled the interplay between charge and superconducting orders~\cite{Lian2023,Yang2018,Simon2024,Hwang2024}. \comment{Even though cuprates and transition metal dichalcogenides exhibit similar features,} The superconducting state there \comment{in the latter systems} is Bardeen-Cooper-Schrieffer-like and can be accounted for by relatively simple mean-field approaches, where the emerging, possibly competing phases result from breaking a particular symmetry~\cite{Marder2010}. For instance, breaking the translation or a point group symmetry promotes a charge density wave (Peierls and Jahn-Teller instabilities); breaking the (spin) rotation symmetry results in a spin density wave, i.e. the material develops magnetism, in a variety of simple flavours (ferromagnets, antiferromagnets) or more exotic spin textures, such as ferrimagnets~\cite{Marder2010}, skyrmions~\cite{Fert2017}, or nematics~\cite{Kim2024} (also see Sec.~\ref{ssec:beyond}).
 
In contrast, the physical mechanism behind high-$T_\mathrm{c}$ superconductivity is intrinsically much more complicated~\cite{Lee2006,Seibold2021} and has been a puzzle for many years, until it was realized that mean-field descriptions are essentially semi-classical approaches missing critical quantum correlations, not only overestimating long-range correlations, and thereby the stability of ordered phases, but also preventing a proper description of materials where entanglement plays a strong role~\cite{Lee2006,Keimer2015,Scalapino2012,Broholm2020}. These considerations helped identify a new phase of matter, broadly named quantum spin liquids~\cite{Balents2010,Savary2016,Norman2016}, which are essentially paramagnetic Mott insulators resulting from the interplay between magnetic and geometric frustration.
In particular, quantum spin liquids are expected to occur on triangular or kagome lattices, and specifically in several families of materials: Herbertsmithite ZnCu$_3$(OH)$_6$Cl$_2$, SrCr$_{9p}$Ga$_{12-9p}$O$_{19}$ or the vdW material $\alpha$-RuCl$_3$, to cite a few~\cite{Norman2016,Sen2011,Broholm2020}.
 
Regarding topological aspects (Sec.~\ref{ssec:topo}), it is nowadays less surprising that many 2D materials host (hidden) topological properties~\cite{Sachdev2003}, along with (non-local) excitations with fractional properties (charge, spin, statistics, etc) associated to underlying gauge theories, such as $\mathbb{Z}_2$, $U(1)$ or Chern-Simons~\cite{Wen2017,Broholm2020}. However, detecting the topological properties and fractionalized excitations in quantum spin liquids remains experimentally challenging. Because entanglement and correlations are central to these systems, they naturally connect to quantum information and quantum computing~\cite{Broholm2020}.

\subsection{Topology of the electronic band structure}
\label{ssec:topo}
\begin{figure}[h!]
\includegraphics[width=75.19mm]{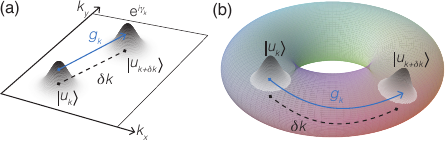}
\caption{Graphical representation of the quantum metric tensor, ${\bf g}_{\bf k}$, and the geometrical phase ${\bm\gamma}_{\bf k}$ between two quantum states at different points in the (a) first Brillouin zone and (b) total Brillouin zone represented as a torus. The metric ${\bf g}_{\bf k}$ is a tensor and generally differs from $\delta{\bf k}$. The color gradient mimics the phase changes occurring along the path ${\bf k}\rightarrow{\bf k}+\delta{\bf k}$. \label{fig:quantumgeometry}}
\end{figure}

Over the past decades, the quantum geometry of wave functions has emerged as a powerful framework for understanding the quantum properties of materials \cite{Provost1980,Resta2011,Torma2023}, driving conceptual advances such as the discovery of ``topological materials'' \cite{Qi2011}. Quantum geometry studies the structural properties of the wave functions $|u_{\bf k}\rangle$ in the space spanned by parameters ${\bf k}$ \cite{Torma2023}, and establishes one-to-one correspondences between physical properties and geometrical concepts, such as metric, connection, and curvature \cite{Ahn2021}. The parameter ${\bf k}$ can be the momentum, time, spin, pseudospin, position, etc \cite{Sundaram1999}. The distance between two wave-functions in ${\bf k}$-space is given by the quantum geometric tensor \cite{Provost1980} 
 \begin{eqnarray}
 {\cal G}_{ij}=\langle \partial_{k_i}u_{\bf k}|\partial_{k_j}u_{\bf k}\rangle-\langle \partial_{k_i}u_{\bf k}|u_{\bf k}\rangle \langle u_{\bf k}|\partial_{k_j}u_{\bf k}\rangle,
 \end{eqnarray}
 where $|\partial_{k_i}u_{\bf k}\rangle$ is the derivative of the quantum state $|u_{\bf k}\rangle$ along direction $i$ of the parameter ${\bf k}$. The real part of this tensor, $g_{ij}= {\rm Re}[{\cal G}_{ij}]$, is called the quantum metric and measures the distance between the states upon a small change $\delta{\bf k}$ in the parameter space (Fig. \ref{fig:quantumgeometry}). The imaginary part $\Omega_l=\epsilon_{ijl} {\rm Im}[{\cal G}_{ij}]$, where $\epsilon_{ijl}$ is the Levi-Civita symbol, is the (Berry) curvature experienced by the quantum state along the path ${\bf k}\rightarrow {\bf k}+\delta{\bf k}$ \cite{Berry1984,Xiao2010b} (see Fig. \ref{fig:quantumgeometry}). This curvature is associated to a geometrical phase, $\delta{\bm\gamma}_{\bf k}=\delta{\bf k}\cdot\bf{\cal A}_{\bf k}$, where the vector potential is directly related to the curvature, ${\bm\Omega}_{\bf k}={\bm\nabla}_{\bf k}\times\bf{\cal A}_{\bf k}$. Several major phenomena in condensed matter have been interpreted in terms of the quantum geometry: the anomalous and spin Hall effects \cite{Nagaosa2010,Sinova2015}, but also pumping \cite{Thouless1983}, spin-orbit torques \cite{Manchon2019}, nonlinear transport \cite{Sodemann2015}, optical excitations \cite{Ahn2021}, and even flatband superconductivity \cite{Peotta2015}.

Quantum geometry has spectacular effects in 2D since the Brillouin zone becomes a torus, over which the integral of Berry curvature defines topological invariants like the Chern number \cite{Qi2011}. These ideas have led to a deep understanding of topologically nontrivial electronic, bosonic, and even interacting many-body states. The emblematic example in 2D magnetic systems is the integer quantum Hall effect, wherein the Hall conductance is quantized in integer multiples of $e^2/h$, corresponding to the Chern number of the filled Landau levels \cite{Thouless1982,Haldane1988}. This quantum \textit{anomalous} Hall effect (QAHE) preludes later discoveries of magnetic topological insulators made of either intrinsic or magnetically-doped vdW multilayers [see, e.g., \cite{Chang2013}]. Importantly, QAHE does not require a homogeneous magnetic configuration: a 2D skyrmion lattice \cite{Hamamoto2015} or a noncollinear magnetic configuration, such as the 3Q triangular antiferromagnet \cite{Shindou2001} supports QAHE.
 
Quantum geometry is acknowledged as the source of (unquantized) nonlinear Hall effect \cite{Sodemann2015,Kaplan2024}. Formally put, at the second order in the electric field, the Hall current reads
\begin{eqnarray}
{\bf j}_\mathrm{H} \propto \tau \int dk {\bf E}\times({\bf E}\cdot\partial_{\bf k}{\bm\Omega})+ \int dk ({\bf E}\cdot\partial_{\bf k}){\bf g_k}\cdot{\bf E}.
\end{eqnarray}
The first term is proportional to the Berry curvature dipole, and is finite only if both mirror and inversion symmetries are broken. It is relevant in nonmagnetic systems, e.g., WTe$_2$ bilayers \cite{Ma2019,Kang2019}. The second term is proportional to the quantum metric, and is nonzero when both {\em time} and inversion symmetry are broken. In other words, it requires magnetism. This contribution has been reported in the MnBi$_2$Te$_4$ vdW antiferromagnet, where the Berry curvature dipole is cancelled by symmetry \cite{Gao2023,Wang2023b}. Recently, the quantum metric has also been proposed to give rise to nonlinear longitudinal transport \cite{Das2023,Kaplan2024} and has been reported experimentally in the MnBi$_2$Te$_4$ \cite{Wang2023b} and CrSBr \cite{Jo2025} vdW antiferromagnets. Quantum geometry concepts have also been extended to magnons, in analogy to topological electronic materials (see Sec.~\ref{ssec:topoMagnons}). Again, the number of theoretical studies is incommensurate with the handful of experiments reporting signatures of magnon topology. Alongside tremendous progress in the realm of topological physics, the realization of intrinsically 2D magnetic topological phases is an ongoing challenge. 

\section{From conventional 2D magnetism to new classes of effects}
\label{sec:whatsnew}

Let us now try and identify unique physical effects explored with 2D magnets, their heterostructures and their parent vdW crystals.

\subsection{Availability of 2D/vdW magnets}
\label{ssec:materials}

Several bulk vdW magnets have been studied starting from the 60s, including transition metal halides and transition metal phosphorous trichalcogenides \cite{Wiedenmann1981}. Years later, truly 2D materials were successfully isolated \cite{Novoselov2005}---not magnetic ones though, at least for more than a decade. \textit{Ab initio} predictions of magnetism in realistic 2D crystals were made first for a perovskite \cite{Sachs2013} and an oxide \cite{Kan2013}, and shortly after, picking up candidate materials already known in their bulk phase, especially trichalcogenides \cite{Sivadas2015} and halides \cite{Zhang2015}. Experimental confirmation of their magnetism followed soon after. While antiferromagnetic order is generally much harder to detect in 2D than in bulk compounds \cite{Wang2016,Lee2016,Scagliotti1985}, 2D ferromagnetism was readily observed in both semiconducting \cite{Gong2017,Huang2017} and metallic \cite{Deng2018} systems. 

In all these reports, and many others subsequently, samples were obtained in the form of micro-flakes using mechanical exfoliation, with special precautions, however (see below). Efforts to grow these materials over large areas from the bottom-up, by molecular beam epitaxy (MBE), were contemporary \cite{Liu2017}, and single-layers rapidly became accessible accordingly \cite{Chen2019}. Deposition techniques (MBE, chemical vapour deposition, etc) offer fine control over the thickness, composition, and purity. Nevertheless, precisely selecting the number of 2D layers or a specific material phase remain delicate matters \cite{Zhou2023}. Phase control and reliable identification are often problematic in the literature, leading to hasty claims of magnetic properties and disturbing disagreements, for instance due to indetermination on compositions, e.g., in Cr$_x$Te$_y$ alloys, having $T_\mathrm{Curie}>$300~K when $x/y=1/2$ \cite{Purbawati2020} but below otherwise \cite{Meng2021,Purbawati2023}, or in Fe$_{5-x}$GeTe$_2$ alloys, in which not-precisely-known $x$ raises questions about the planar or perpendicular magnetic anisotropy \cite{May2019,Yang2020,Ribeiro2022} (strain effects might here play a role too). Careful cross-characterizations are crucial here to ascertain the claims.

Besides, much like most 2D materials beyond graphene, the grown 2D magnets are polycrystalline, with rotated crystallites of size of the order of 100~nm at best [with some exceptions \cite{Prabhu2025}]. This is much smaller than the few-10~$\mu$m-large single-crystals obtained by mechanical exfoliation. Polycrystallinity excludes addressing physical effects rooted in a well-defined lattice orientation, either in real space with moir\'{e} / stacking effects, or in reciprocal space with specific features in a spin-polarized electronic band structure. Additionally, with polycrystallinity come defects (grain boundaries) that might alter, e.g., a ferromagnet's coercivity \cite{Aharoni1960,Kronmueller1987}. Finally, in the case of in-plane magnetic anisotropy, polycrystallinity might yield an apparent planar magnetization isotropy, not to be confused with a (microscopic) XY spin character.

An unfortunate characteristic of most 2D magnets is their transformation in air or under the influence of moisture and light. This is well known for CrI$_3$ layers, which decompose in a few seconds when exposed to both air and light \cite{Shcherbakov2018}. More stable 2D magnets have been identified, although initially often with rather optimistic claims about long term stability \cite{Liu2018c,Son2021,Kim2021,Ye2022}. One such compounds is CrSBr, which is altered at time scales of weeks \cite{Klein2024}. Even if a material does not fully degrade, it may undergo phase transitions that alter its magnetic properties. For instance, CrTe$_2$ transforms into Cr$_x$Te$_y$ under moderate illumination \cite{Riccardi2025}. In Fe${_3}$GeTe$_2$, the top layer may oxidize, turning antiferromagnetic and coupling with the inner ferromagnetic layers \cite{Zhao2023}, reminiscent of the exchange anisotropy first identified in ferromagnetic cobalt particles coated with their own antiferromagnetic native oxide \cite{Meiklejohn1956}.

Protecting sensitive 2D magnets is often required. However, the choice of the capping material is critical. For example, capping with 3D materials may alter the magnetic properties of the underlying 2D layer due to covalent bonding \cite{Gish2021,Galbiati2020}. Capping with another 2D material, such as $h$-BN or graphene, can avoid this issue while also providing the added benefit of high in-plane thermal conductivity. This property can reduce the sensitivity of 2D magnets to intense laser light \cite{Riccardi2025}.

Nonetheless, exfoliation and transfer processes can introduce local strain and interlayer sliding, potentially leading to variations in the magnetic ordering of layered compounds that are sensitive to their stacking arrangement, such as CrI$_3$ \cite{Thiel2019,Ubrig2019}.

\subsection{Predictions for vdW and 2D magnets}
\label{ssec:predMat}



In Sec.~\ref{ssec:micro}, we discussed the difficulties inherent to first principles modeling of magnetic exchange. On top of these are difficulties in properly computing the magnetic properties, including the ordering temperature. This is usually estimated using Monte Carlo simulations on an effective spin Hamiltonian accounting for various forms of exchange interactions \cite{Evans2014}. Whereas bilinear terms (Heisenberg exchange and magnetic anisotropy) are easy to compute from first principles, most simulations neglect higher-order terms [e.g., biquadratic and directionally-dependent exchange \cite{Jackeli2009}]. These terms, usually negligible in bulk 3D systems, can be large in 2D/vdW materials such as RuCl$_3$ \cite{Plumb2014}. Yet, only a few studies have examined them \cite{Xu2018,Kartsev2020}. Since an accurate Hamiltonian is essential for reliable $T_{\rm Curie}$ predictions, early works relying on an Ising approximation (infinite anisotropy) substantially overestimated the critical temperature \cite{Wang2022}.

\begin{figure}[h!]
\includegraphics[width=8cm]{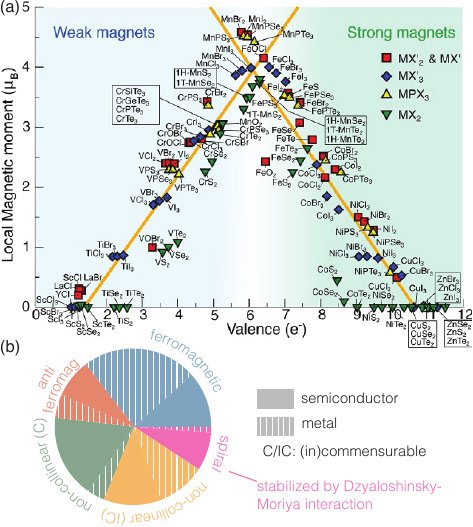}
\caption{(a) Local magnetic moment at the metal atom as a function of its valence. From~\cite{Wang2022}. (b) Distribution of magnetic orders of 154 materials with nonzero magnetic moments in the computational 2D materials database (C2DB). Adapted from~\cite{Sodequist2024}.\label{figDFT}}
\end{figure}

Early theoretical investigations of 2D/vdW magnets, XI$_3$ \cite{Zhang2015} and MPX$_3$ \cite{Chittari2016} compounds, used hybrid functionals for non-local exchange combined with a Monte Carlo estimate of the transition temperature. Different approaches were carefully compared \cite{Chittari2016}, pointing out the high sensitivity of the results to the choice of functional, and raising questions about the predictive power of first-principles methodologies. Multiple subsequent predictions of 2D/vdW magnets followed, based on high-throughput (HT) calculations (with potentially the same pitfalls). Such calculations use the chemical composition and crystal structure as input, usually obtained from known databases \cite{Haastrup2018,Zhang2021}. The key point is selecting a simple descriptor---such as the band-gap, formation energy, or magnetic state---to organize results. To accelerate calculations, DFT+U is often used, since hybrid functionals are too expensive and, to date, unfit for HT simulations. Despite this shortcoming, HT simulations can reveal trends, such as the effect of composition on magnetic properties, as exemplified by Fig.~\ref{figDFT}(a) \cite{Wang2022}. HT approaches combining DFT+U and Monte Carlo methods have also identified potential 2D magnets with above-room-temperature ordering \cite{Kabiraj2020,Torelli2020}. Although HT predictions are not quantitatively reliable, they provide valuable chemical trends that can guide experiments and motivate targeted hybrid-functional validation.

The discovery of materials can be accelerated by machine learning and data mining approaches, which reduces computational cost by learning from a relatively `small' dataset. A first difficulty is the scarcity of the training data (few 1,000 data points at best), as only a tiny portion of the explored structures turns out to be both stable and magnetic. To illustrate this, among $>$10,000 candidates derived from 240 compounds with a 2$\times$MnBi$_2$Te$_4$ structure, only 25 met criteria for stability, large magnetic moment ($>4.55$~$\mu_B$) and a positive band-gap \cite{Bhattarai2023}, with just nine predicted to be ferromagnetic insulators. Similar methods have been used to identify 2D materials with large orbital magnetic moment \cite{Ovesen2024} or noncollinear magnetic ground state \cite{Sodequist2024} [Fig.~\ref{figDFT}(b)].

Artificial intelligence algorithms represent an extraordinary opportunity for the search for new materials with designer properties. Nonetheless, the \textit{ab initio} methodology (DFT+U, assumed collinear magnetism) used in the vast majority of studies so far calls for caution. More precise functionals are required to increase the credibility of predictions. That being said, the obtained results are already very encouraging, and several new classes of materials have already been proposed. The recent investigation of heterostructures opens additional perspectives \cite{Pakdel2024,Bhattarai2025}.

\subsection{Dimensionality crossover}
\label{ssec:2Dmag}

The existence or absence of order and phase transition in 2D has been, and remains today, a central question in magnetism. It has been addressed both theoretically (Sec.~\ref{ssec:stat}) and experimentally in thin films (Sec.~\ref{ssec:micronano}) and bulk vdW crystals (Sec.~\ref{ssec:historicalmagnon}) since the second half of the 20th century. Remarkably, the motivations driving the study of vdW magnets today echo those of surface scientists from that era: ``Does ferromagnetism exist in films as thin as 1 or 2 monolayers? What does the phase transition look like? What are the values of the critical exponents?'' \cite{Rau1989}.

2D/vdW materials are often expected to be only weakly influenced by their environment, raising hopes that they could help resolve those long-standing questions. Recent experiments, however, suggest that this expectation is too optimistic. A key observation is the general decrease of $T_\mathrm{Curie}$ with thickness in vdW magnets \cite{Gong2017}, a trend reminiscent of conventional materials and captured by statistical mechanics models addressing the role of boundary conditions \cite{Fisher1972}. A quantitatively similar thickness-dependent decay in vdW magnets, despite their very small interlayer exchange, may seem surprising, and raises questions about the role of other effects, e.g., structural details (hence electronic and magnetic ones) tied to interlayer vdW forces.

Back to critical exponents, $\beta$ was first determined in a 2D magnet with Fe$_3$GeTe$_2$ \cite{Fei2018}, wherein a crossover was found from a 2D-Ising-like value for the monolayer to higher values for thicker flakes, better corresponding to the 3D Ising case. Going beyond the sole estimate of a single critical exponent, several detailed analyses determined two of them. In bulk Fe$_3$GaTe$_2$\comment{ (iso-structural to the above mentioned Fe$_3$GeTe$_2$)}, the exponent values match a 3D Heisenberg model \cite{Algaidi2024}, while for bulk CrI$_3$, conclusions were more puzzling, showing no clear resemblance with a single universality class \cite{Lin2018,Liu2018}. In MPX$_3$ compounds, thickness reduction was found to have either no effect, with FePS$_3$ having a $\beta$ exponent typical of a 2D XY model already in the bulk form, or, on the contrary, to come along with a dimensionality crossover, with $\beta$ in CoPS$_3$ decreasing progressively from a 3D Ising to a 2D XY value \cite{Houmes2023}. A 2D XY model was also found to account for $\beta$ in single-layer CrCl$_3$ over a small temperature window \cite{BedoyaPinto2021}, while inferring critical exponents reliably is usually based on scaling-law analyses over several orders of magnitudes of reduced temperature \cite{Back1995}. Overall, determining universality classes remains a difficult task experimentally, and interest in vdW magnets remains high for testing theoretical predictions.

\subsection{Tuning magnetic properties}
\label{ssec:tuning}

There are multiple ways to manipulate the magnetism of 2D/vdW crystals. These include conventional methods such as tuning the composition or applying strain, as well as approaches that exploit the atomically-clean interfaces of vdW materials, the extreme thinness of 2D magnets, and the absence of covalent bonding when assembling (vdW) heterostructures. Such characteristics are valuable for combining various properties and exploiting electric field or proximity effects.

\subsubsection{Composition}
\label{ssec:compo}

Except for graphene, which can become magnetic by proximity (Sec.~\ref{sssec:proximity}) or when it undergoes a quantum phase transition induced by strong electronic correlations (Sec.~\ref{ssec:beyond}), 2D/vdW magnets are always alloys. Their properties are controlled by chemical composition. In traditional materials, this is the way electronic band gaps are tuned, and also the route to engineering Curie, N\'{e}el, or compensation (between magnetic sublattices) temperatures in ferromagnets, antiferromagnets, or ferrimagnets, respectively---researchers even use phenomenological abacuses \cite{Graf2011,Castelliz1955,Kanomata1987,Lee1975}. Adjusting the composition also gives access to magnetic quantum phase transitions, towards e.g., spin density waves or antiferromagnetic states, as shown for example in CeCu$_{6-x}$Au$_x$ \cite{Schroeder2000}, Cr$_{1-x}$V$_x$ \cite{Yeh2002} or Nb$_{1-x}$Fe$_{2+x}$ \cite{Moroni2009}.

However, playing with the composition has not been a mainstream option to control magnetism of 2D/vdW materials, because the widespread use of mechanically exfoliated samples makes stoichiometric variation impractical. An alternative is to grow samples from the bottom up with the desired composition. Interestingly, the change of structure (compression/expansion of the lattice) related to a composition change is potentially not as problematic as it is with traditional materials, where lattice mismatches spawn interface defects coming with strain fields or extending throughout the layers. In contrast, there are no strong chemical bonds at interfaces between vdW layers, and moir\'{e} patterns accommodate lattice mismatches.

The coarsest form of composition tuning is a change of structural phase. There, the atomic coordination is modified, as is the case from Fe$_3$GeTe$_2$ to Fe$_4$GeTe$_2$ and Fe$_5$GeTe$_2$ \cite{May2019,Seo2020} or from FeS to FeS$_2$ \cite{Zhou2023}. $T_\mathrm{Curie}$ can be tuned accordingly, in some cases very close to room temperature. A finer tuning consists of substituting one of the atoms. This is achieved by supplying less of one element and more of another during growth. Otherwise, the excess material tends to intercalate between the layers, creating strong bridges between them \cite{Fujisawa2020,Guan2023}, and the material is not a vdW one anymore. Super-exchange interactions between the magnetic ions (see Sec.~\ref{ssec:micro}) imply that both the magnetic and non-magnetic ions play a role in magnetic order. Substituting the magnetic ion allowed for increasing $T_\mathrm{Curie}$, switching the magnetic anisotropy direction, even the kind of order (ferro/antiferromagnetic) in Fe$_{5-x}$Co$_x$GeTe$_2$ alloys \cite{Tian2020,May2020}, and changing the magnon gap in Fe$_{1-x}$Ni$_x$PS$_3$ \cite{LeMardele2024}. Substituting the chalcogen in MnPS$_x$Se$_{3-x}$ compounds proved to alter the interlayer and intralayer spin interactions, especially the spins' orientation, which was interpreted as an effect of `internal chemical pressure' [anions of different kinds, here S or Se, locally expand/compress the lattice, as demonstrated previously with more traditional materials \cite{Aeppli1997}] \cite{Baral2024}. Another example is the strong $T_\mathrm{Curie}$ enhancement upon substitution of Ge with Ga in Fe$_3$GeTe$_2$, achieved either by growth in the presence of Ga \cite{Zhang2022} or Ga$^+$ implantation \cite{Yang2020}. The latter may be more or less intentional when patterning magnetic micro/nanostructures with focused ion beams, as reported decades ago \cite{Chappert1998,Rettner2002}.


Substitution in minute amounts is a route to turn traditional non-magnetic III-V and II-VI semiconductors into ferromagnets whose $T_\mathrm{Curie}$ and anisotropy could be tuned with an electric field \cite{Chiba2006,Chiba2008}. Stabilizing such diluted magnetic semiconductors with, e.g., Mn (a high-spin ion) has proved to be very challenging due to the tendency to spinodal decomposition, wherein magnetic ions tend to cluster rather than evenly substitute the group-III or -II element \cite{Dietl2015}. The challenge presumably remains with vdW materials: local strains associated with the atomic substitution might be less critical (accommodated by affordable variations of the vdW gap), but intercalation might be more favorable than substitution. Magnetism is predicted in Mn-substituted MoS$_2$, via exchange interactions between spins localized around the Mn ions \cite{Mishra2013}. While some experimental reports suggest successful growth of diluted ferromagnets [Mn:MoSe$_2$ layers \cite{Dau2019}], others highlight issues like clustering of the magnetic element \cite{Gay2021}. The long-standing debate about clear-cut evidence of magnetism induced by dilute substitution (when it is successful) is revived with 2D materials \cite{Mallet2020}, and past works on traditional materials prescribe caution, especially when exciting conclusions are drawn from a single kind of experimental probe revealing unconventional signatures \cite{Yun2020}. It is worth noting that dilute substitution is one approach to turn a semiconductor into a magnet, but another potential method involves exploiting proximity effects (see \ref{sssec:proximity}).

While full disorder is targeted in diluted systems, the equiatomic substitution raises the question of the spatial distribution of the atomic elements. Indeed, a single layer consists of atomic sublayers, and the alloying atoms (say S/Se in a chalcogenide, Br/I in a halide) may be located in distinct layers. Such Janus structures are expected to come with a built-in vertical electrical polarization that shall alter superexchange coupling between magnetic ions \cite{Cui2020,Xu2020,Yuan2020}, and lack inversion symmetry (unlike most pure vdW compounds). The latter authorizes a DMI between magnetic ions, with a magnitude sometimes comparable with that in Co/Pt or Fe/Ir layers, as calculated for transition metal sulfide+selenide \cite{Liang2020}. Janus vdW magnetic alloys, which have not been reported experimentally yet, may allow for stabilizing noncollinear spin textures, QAHE, or polarizing the valleys in the electronic band structure in 2D semiconductors \cite{Jiang2023}.

\begin{figure}[h!]
\includegraphics[width=8cm]{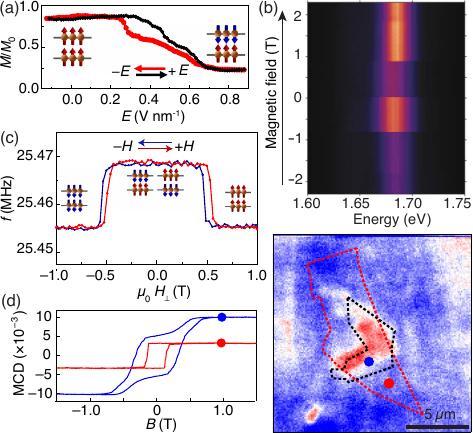}
\caption{(a) Electrical switching of magnetic order in bilayer CrI$_3$, observed in the magnetization (normalized to $M_\mathrm{s}$) vs electric field ($\mu_0 H$ = 0.44~T). Adapted from \cite{Jiang2018a}. (b) Changes in photoluminescence spectra (left-circular polarization) of single-layer WSe$_2$, induced by changes of interlayer magnetic order in a thin CrI$_3$ underneath WSe$_2$, controlled by magnetic field. Adapted from \cite{Zhong2017}. (c) Resonance frequency of a suspended CrI$_3$ bilayer membrane (sandwiched between graphene and WSe$_2$), as function of applied magnetic field (and different interlayer magnetic orders). Adapted from \cite{Jiang2020}. (d) Ferromagnetic and antiferromagnetic stackings in twisted bilayer CrI$_3$, observed with magnetic circular dichroism. The map measures to dichroic signal at 1~T, relative to the one at 0~T. The edges of the two layers are highlighted. Adapted from \cite{Xu2022}. \label{fig:tuning}}
\end{figure}

\subsubsection{Electric-field effects}

Using electric fields, instead of currents and their related resistive heat losses, holds promise for energy-efficient manipulation of magnetism. The effects of electric fields on metals are, however, restricted to their surface, and are therefore most efficient with ultra-thin films with which they were demonstrated in a variety of flavours \cite{Weisheit2007,Maruyama2009,Matsukura2015}. Diluted magnetic semiconductors (see previous section) are naturally sensitive to electric fields even beyond the 2D limit. Obviously, 2D magnets, with their ultimate thickness, are promising alternatives. When stacked with other vdW materials, they additionally form interfaces free of interfacial defects such as dislocations. {Integrating} 2D magnets within dual-gate devices, a kind of architecture that emerged with graphene \cite{Oostinga2008}, the influence of an electric field and of the charge carrier density can even be investigated separately, or combined. Magnetism can thereby be influenced electrically through mechanisms like linear magnetoelectric coupling or electrostatic doping.

Linear magnetoelectric coupling occurs when a material lacks both time-reversal and inversion symmetries---a condition for instance satisfied in bilayer CrI$_3$ in its antiferromagnetic ground state, but not in its ferromagnetic phase or as a monolayer. It was found that magnetoelectric coupling peaked near the spin-flip transition ($\sim$0.5~T), enabling electric switching between antiferromagnetic and ferromagnetic phases at a constant magnetic field (Fig.~\ref{fig:tuning}a). This effect originates from the electric-field dependence of the interlayer exchange bias, especially in vdW bilayers owing to their electrically-tunable layer asymmetry. \cite{Jiang2018a}.

Tuning the charge carrier density, via a gate dielectric between an electrode and the 2D material is another route for modulating magnetism, without the above-mentioned broken symmetry requirement. The technique was first applied to bilayer \cite{Jiang2018b,Huang2018} and monolayer \cite{Jiang2018a} CrI$_3$, in a dual-gate setup operating free of the influence of an electric field. In the monolayer, properties like $M_\mathrm{s}$, $T_\mathrm{Curie}$, and coercive field ($H_\mathrm{c}$) increase (decrease) with hole (electron) density \cite{Jiang2018a}. In the bilayer, electron doping of a few 10$^{13}$~cm$^{-2}$ drastically reduces the stability range of the antiferromagnetic phase, favoring the ferromagnetic phase \cite{Jiang2018b}. All-electrical fully-reversible switching was not demonstrated back then, yet these findings represent a significant progress in this direction \cite{Jiang2018b,Huang2018}.

Instead of a solid-state dielectric, an ionic liquid may be advantageous to reach high charge carrier concentrations in excess of 10$^{13}$~cm$^{-2}$ \cite{Dhoot2006,Panzer2006} and hopefully demonstrate drastic changes of magnetic properties. In multilayer CrGeTe$_3$, both electron and hole doping achieved accordingly reduced the saturation field and increased $M_\mathrm{s}$, while $H_\mathrm{c}$ and $T_\mathrm{Curie}$ were largely unaffected \cite{Wang2018b}. This behavior may result from a redistribution of magnetic moments due to changes in the spin-polarized band structure although a quantitative understanding is hindered by the yet unknown, material-dependent carrier density. In contrast, a significant $T_\mathrm{Curie}$ rise was reported, from 60 to 200~K \cite{Verzhbitskiy2020}. A striking switch of magnetic anisotropy from out-of-plane to in-plane was also found, attributed to a carrier-mediated double-exchange mechanism overtaking the intrinsic superexchange interaction of the insulating state.

Voltage-control of magnetism was reported for a \textit{metallic} vdW magnet too, Fe$_3$GeTe$_2$ (CrI$_3$ and CrGeTe$_3$ are semiconductors). In a trilayer, ionic gating was claimed to raise $T_\mathrm{Curie}$ from $\sim$100~K to near room temperature, and $H_\mathrm{c}$ was found to vary in tandem with $T_\mathrm{Curie}$ \cite{Deng2018}. This significant enhancement was argued to stem from a high level of electron doping (claimed to be $\sim$10$^{14}$ cm$^{-2}$ per layer).

Ionic liquid gating, however, may come with side-effects, and in particular intercalation of the gating medium between layers, chemical changes, and even material decomposition \cite{Weber2019}. The consequences on the lattice or electronic structure, magnetic state of individual ions and magnetic interactions (and $T_\mathrm{Curie}$) are then expectedly irreversible and go beyond the intended electrostatic tuning. Distinguishing pure electrostatic effects from material modifications is non-trivial and often requires advanced physico-chemical characterizations, sometimes performed \textit{in situ}.

A third, increasingly popular strategy to electrically control magnetism in 2D materials involves interfacing them with ferroelectric layers, enabling non-volatile electrical control of magnetism that persists after the electric field is removed. 
This was first demonstrated using a thin ferroelectric polymer layer deposited on top of Cr$_2$Ge$_2$Te$_6$ \cite{Liang2023}. Applying $\pm$5~V across this heterostructure modulated the magnetic hysteresis loop, effectively toggling the magnetization between two stable states without continuous power input. Notably, the modulation was observed with bi- to four-layers, but not in thicker samples, underscoring the role of interfacial effects and the emergence of multiferroic behavior at the nanoscale. Pushing this concept further, all-vdW heterostructures were assembled, pairing ferromagnetic Fe$_{3-x}$GeTe$_2$ with the ferroelectric semiconductor In$_2$Se$_3$  \cite{Eom2023}, demonstrating a significant reduction of $H_\mathrm{c}$ under both gate-voltage polarities, which was found to correlate with a voltage-induced in-plane lattice expansion in both materials. Although these findings illustrate a promising pathway to low-power, non-volatile spintronic devices, several challenges remain, especially the limited availability of suitable ferromagnetic and ferroelectric vdW materials and the precise control and reproducibility of interfacial coupling.

\subsubsection{Proximity effects}
\label{sssec:proximity}
Any pair of 2D materials can, in principle, be stacked into vdW heterostructures, enabling combinations of properties that are difficult to achieve in conventional materials, such as ferromagnets coupled to semiconductors, superconductors, topological insulators, or strong spin–orbit systems.
Not only two functionalities can be combined, but also a mutual control can be achieved. The electronic or spin properties of one layer then are modified by the neighbor layer, in an at-equilibrium fashion (Fig.~\ref{fig:timeline}). This typically occurs at the nanometer scale through wave-function overlap, effectively creating a new interfacial material. Although the proximity effect is an equilibrium phenomenon, it is often probed by non-equilibrium processes---e.g., stray fields, spin injection, or charge transfer---, which has led to confusion and occasional misuse of the term.


First identified in superconductors \cite{Holm1932}, the proximity effect was later explored in non-magnetic metals adjacent to ferromagnets \cite{Hauser1969}. In ferromagnet/semiconductor hybrids, it consists of a spin polarization of semiconductor carriers via interfacial exchange, enabling reciprocal sensing of magnetization without external fields. When the interfacial density of spin-polarized carriers matches that of magnetic ions, the system behaves as a single spin ensemble whose magnetic response can be tuned optically or electrically \cite{Zakharchenya2005}. Despite decades of experimental efforts using conventional systems \cite{Myers2004,Korenev2012,Korenev2016}, the observed proximity effects remain generally weak, mainly due to imperfect interfaces and thick ferromagnetic layers wherein interfaces contribute only marginally.

This is precisely why 2D ferromagnet/semiconductor vdW heterostructures are appealing. A first experimental demonstration was obtained with a semiconducting WSe$_2$ monolayer on a few layers of the vdW magnet CrI$_3$ \cite{Zhong2017}. Below the $T_\mathrm{Curie}$ of CrI$_3$, WSe$_2$ photoluminescence becomes circularly polarized without an external field (Fig.~\ref{fig:tuning}b), although this alone could arise from non-equilibrium spin-dependent charge transfer under laser excitation. The unambiguous signature came from the Zeeman splitting of  opposite-helicity photoluminescence peaks, from which the known exciton Land\'{e} $g$ factor yields a proximity-induced exchange field of approximately 13 T. Although similar effects have been widely reported, only a few have directly demonstrated Zeeman splitting \cite{Ciorciaro2020,Zhang2020,Shayan2019}.

{Despite high expectations, the research on 2D ferromagnet/semiconductor vdW heterostructures has progressed rather slowly, partly due to the limited number of material combinations with clear proximity effects.}
Insulating magnets such as CrX$_3$ or CrSBr are typically preferred over metallic ones as the latter tend to quench the optical properties of 2D semiconductors. Another constraint is the requirement for ultraclean, atomically sharp interfaces to observe the short-range proximity effect, demanding high-quality fabrication techniques.

Only few studies have explored the inverse proximity effect, i.e., using the semiconductor to manipulate the magnetic properties. The magnetic hysteresis loop of CrI$_3$, forming a heterostructure with WSe$_2$, was found to be modified under optical excitation, but laser-induced heating could not be ruled out as the cause \cite{Seyler2018}. More recently, magnetic domain formation and switching were optically induced using femtosecond pulses in the same kind of heterostructure \cite{Dabrowski2022}. However, the underlying mechanism remains unclear and may involve interfacial charge transfer rather than exchange-driven proximity interactions.

{Besides 2D semiconductors, magnetic exchange coupling was also induced in (semimetallic) graphene adjacent to a 2D magnetic insulator to enable spintronic functionalities while preserving its high mobility} (minimal chemical doping or structural damage). Graphene, placed on the interlayer antiferromagnet CrSBr, exhibits proximity-induced spin polarization that can be then used to inject spin currents, both electrically and thermally, as well as the anomalous Hall effect \cite{Ghiasi2021}. Such a spin polarization {has been reported to be tunable by electrostatic gating across the Dirac point} \cite{Yang2024b}{, although further work is needed to assess the robustness and reproducibility of this effect}. Seamless, all-graphene spin valves were further obtained by selectively proximitizing certain graphene regions with the ferromagnetic insulator CrGeTe$_3$, a heterostructure that also displayed spin Hall and anomalous Hall effects owing to additional spin-orbit proximity \cite{Yang2025}. Spin-orbit and magnetic exchange couplings were also achieved in graphene in proximity with CrPS$_4$ (another interlayer antiferromagnet) to demonstrate a coexistence of quantum spin Hall effect and magnetism in graphene \cite{Ghiasi2025}. These results illustrate how proximity to 2D magnets can {induce spin polarization and spin–orbit coupling in graphene, offering a promising route toward engineering magnetic and topological phases, although their robustness and scalability remain open challenges}.

{Finally, proximity to a dielectric modulates the electrostatic screening and can thereby influence the electronic properties of a vdW material \cite{Raja2017,vanLoon2023}. Since electron-electron Coulomb interactions control the magnetic exchange interaction, it was recently suggested that the magnetic ground state of 1$T$-TaS$_2$ could be tuned by proximity \cite{Chen2022b}. To the best of our knowledge, this idea has not been experimentally tested on 2D/vdW magnets yet.}

\subsubsection{Stress effects, including pressure effects}
\label{sssec:stress}

The coupling between structural and magnetic orders yields striking macroscopic effects, first noted by Joule and Villari in the 19th century, and can be described phenomenologically by relating magnetization orientation to strain and stress tensors, leading to a magneto-elastic energy \cite{Kittel1949,Lee1955}. At the microscopic scale, this energy {is tied to the spin-orbit interactions within the atoms,} and has different possible origins: the single-ion anisotropy, altered by a change of crystal-field around the magnetic ions, or the exchange interactions as the bond lengths/angles are modified \cite{Helman1968} (see Fig.~\ref{fig:timeline}). This can affect $T_\mathrm{Curie}$ and magnetic anisotropy in 3D ionic magnets where indirect exchange dominates, e.g., in LaMnO$_3$ \cite{Hwang1995}. These effects are tuned through ``chemical pressure'' (substitution), using a pressure cell \cite{Hwang1995} (kPa to a few 100~GPa) or heteroepitaxial strain in thin films \cite{Hwang1995,Tsui2000}. In traditional materials, however, film roughness and interfacial alloying can diminish magneto-elastic responses \cite{Sander1999}.

These thin-film effects are expectedly irrelevant for 2D/vdW magnets, whose surface is atomically flat, and whose interaction with the substrate (and interfacial stress) can be very small. Interfacial stress is even nonexistent in suspended 2D materials, and mechanical resonators can be assembled in this configuration. The 2D drum's mechanical resonance frequency then depends on the amplitude of the applied magnetic field, which changes the magnetic order between successive layers in CrI$_3$ (Fig.~\ref{fig:tuning}c), while reciprocally $H_\mathrm{c}$ can be tuned with the applied stress \cite{Jiang2020}. An abrupt change in the drum's mechanical properties is also apparent when crossing the magnetic order temperature \cite{Siskins2020,Houmes2023}.

The lamellar nature of vdW materials implies a very peculiar response to applied stress. A hydrostatic pressure, produced by a piston or a diamond anvil cell for instance, produces very anisotropic strains. The materials deform typically 10 times more perpendicular to layers than in their planes, with most of the stress accommodated by a reduction of the vdW gap between layers \cite{Telford2023}, and to some extent as well (Fig.~\ref{fig:timeline}), by a change of bond angle/length \cite{Bykov2013,Haines2018}. The latter results in a modification of the intralayer indirect exchange spin interactions, as shown in Fe$_3$GeTe$_2$ \cite{Ding2021}, while the former changes the interlayer spin interactions in CrSBr, for instance \cite{Telford2023,Pawbake2023}, overall altering the ordering temperatures and more generally the magnetic phase diagrams. The magnetic phase diagrams can be complex, featuring strong magneto-elastic effects such as found across structural phase transitions in FePS$_3$, FePSe$_3$ \cite{Wang2018,Haines2018,Coak2021} or CrOCl \cite{Schaller2023}.

The weak vdW interaction additionally allows consecutive layers to relatively slide under stress (Fig.~\ref{fig:timeline}), {which has no equivalent in other materials, and is accompanied by a change in the interlayer spin interactions}. In CrI$_3$, such a structural transition occurs irreversibly at increasing pressure, and the interlayer magnetic order switches from antiferromagnetic to ferromagnetic \cite{Song2019,Li2019}.

While pressure cells induce \textit{compressive} strain, \textit{tensile} strain can be achieved using stretching devices and may also arise, {in a more-or-less controlled way, in bubbles (or domes) formed following various strategies \cite{Yildirim2025}. Bubble-related} strain already proved to pin domain walls and alter $M_\mathrm{s}$ and $H_\mathrm{c}$ \cite{Grebenchuk2024}. Transferring 2D/vdW materials onto a variety of substrates offers several options. Stretching or bending polymer substrates \cite{Wang2020c,Zhang2022b,Diederich2022}, and more elaborate devices based on rigid gapped substrates \cite{Cenker2022,Du2023}, can be used to apply uniaxial strains (whose exact magnitude in the few 0.1 to few 1\% range depends on the potential sliding of the to-be-strained material on its substrate). This was used to enhance $H_\mathrm{c}$ in Fe$_3$GeTe$_2$ or VI$_3$ \cite{Wang2020c,Zhang2022b}, to modify magnetic domains in Cr$_{1/3}$TaS$_2$ \cite{Du2023}, to control the interlayer spin interaction \cite{Cenker2022} and tune the excitons-magnons coupling in CrSBr \cite{Diederich2022}. Finally, the contact with a substrate can also be a source of strains. Thus, shaping a vdW magnet into a ribbon and bending the ribbon laterally (something that seems hard to achieve with more traditional materials) was shown to generate inhomogeneous $H_\mathrm{c}$ distributions \cite{Bagani2024}.


Strikingly, most efforts have focused so far on relatively thick vdW materials, beyond the few-layer limit, apart from a few exceptions, {where membrane resonators and electronic devices brought sufficient sensitivity} \cite{Song2019,Jiang2020,Houmes2023}.

\subsubsection{Stacking and moir\'{e} effects}
\label{ssec:stacking}

{The amenability of 2D materials to stacking almost at will, with control on the twist angle between adjacent layers, has revolutionized solid-state physics since 2017. Stacking yields moir\'{e} superlattices, whose role as electrostatic superpotentials has been exploited to engineer a multitude of intriguing correlated electronic phases or lateral confinement effects. The concept of twisted 2D magnets, introduced in 2018, promises spatially-varying magnetic textures and nontrivial spin configurations \cite{Sivadas2018,Tong2018} with no equivalent in conventional magnets.}

Most experimental studies have focused on CrI$_3$ twisted layers. {There, the magnetic interaction varies from ferromagnetic to antiferromagnetic depending on the local stacking configuration, i.e., the position within the moiré, whose period reaches tens of nanometers for small twist angles} (Fig.~\ref{fig:tuning}d). {A non-collinear spin configuration is expected with, at certain moir\'{e} sites, spins pointing in different directions in each layer} \cite{Hejazi2020}. A clear experimental demonstration {has been} elusive, {as magnetic characterizations \cite{Xie2022,Xu2022,Xie2023,Yang2024} have so far been limited to micrometer scale, well beyond the moir\'{e} period. Notably, these studies often neglect moiré inhomogeneities due to lattice relaxation at very small twist angles.}
Lattice distortions across moir\'{e}s {(``reconstructions'') are by now suspected to alter, e.g., the excitonic spectra \cite{Andersen2021,Tran2021}}, and their effect as regards moir\'{e} magnetism has been pointed out more recently in CrBr$_3$ devices \cite{Yao2024}. These multiple, often intertwined sources of unconventional behavior underscore the need for {realistic} theoretical modeling, {based on careful descriptions of structural details and of interlayer magnetic interactions (unrealistic numbers have been considered in several works). At the experimental level, high-resolution magnetic probes seem indispensable. Scanning color-center magnetometry ($\sim$50~nm resolution) for instance allowed to distinguish ferromagnetic and antiferromagnetic regions in twisted CrI$_3$ moir\'{e}s \cite{Li2024,Song2021}, but other imaging techniques should be relevant. Besides,} non-collinear magnetism has not been conclusively confirmed, and disorder has hindered the clear observation of the periodicity. Identifying signatures of moir\'{e} magnons \cite{Ganguli2023}, spin-polarized scanning tunneling microscopy, X-ray magnetic circular dichroism, possibly in combination with photoelectron emission microscopy, may provide further insights in the future.

CrSBr, with its intralayer in-plane magnetization and antiferromagnetic interlayer coupling, is another promising material for tuning magnetic properties through stacking and twist, {especially the magnetoresistance in tunnel junctions, whose hysteresis can be field-controlled \cite{BoixConstant2024,Healey2024}, and even nonvolatility can be obtained for certain values of the twist angle \cite{Chen2024}.}

Importantly, several of these magnetic states can be tuned not only by twist angle and temperature but also through electrical gating \cite{Xu2022,Cheng2023}, providing exciting opportunities for functional control in {spintronic and quantum devices} (Sec.~\ref{ssec:transport}). Moreover, several studies predicted that magnetization could be coupled to a spontaneous electric polarization in twisted or properly aligned 2D magnetic layers, potentially leading to multiferroic ordering induced by twisting. For example, a spontaneous electric polarization driven by a non-collinear spin texture may appear in CrX$_3$ twisted layers, resulting from spin-orbit coupling \cite{Fumega2023}. Additionally, the predicted coexistence of magnetic orders with sliding ferroelectricity \cite{Bennett2024} has recently been explored experimentally \cite{Fox2025}.

{Hopefully, high-resolution experimental data will soon allow for testing the abundant theoretical predictions.}

\subsection{Spin excitations}

\subsubsection{Detection of spin excitations in single-layers}

Magnon energies typically fall in the gigahertz (1~GHz $\sim$ 4~$\mu$eV $\sim$ 0.03~cm$^{-1}$) and terahertz (1~THz $\sim$ 4~meV $\sim$ 33~cm$^{-1}$) ranges, with the former corresponding to ferromagnets and their weak anisotropy field, and the latter to antiferromagnets and their strong exchange interaction between spin sublattices \cite{Pashkin2013}. Various experimental techniques can detect and characterize magnons in bulk vdW magnets, including scattering methods---most notably inelastic neutron scattering---and resonance-based approaches such as ferromagnetic resonance, microwave absorption, and terahertz spectroscopy. For example, magnon band structures have been determined by inelastic neutron scattering in (bulk) CrSiTe$_3$ \cite{Williams2015}, CoPS$_3$ \cite{Kim2020}, and MnPSe$_3$ \cite{Liao2024}. Excitations beyond linear magnons have been identified in FeI$_2$ \cite{Bai2023}, CoI$_2$ \cite{Kim2023} and YbCl$_3$ \cite{Sala2021,Sala2023}. However, these techniques lack sensitivity to probe truly 2D magnets, especially exfoliated flakes whose lateral dimensions are typically smaller than the footprint of the probing beam. Most highly space-resolved techniques are currently only sensitive enough for thin films with thickness of the order of few 10~nm, such as electron energy loss spectroscopy in a transmission electron microscope [where strong phonon contributions typically obscure magnon signals \cite{Kepaptsoglou2025}], scanning electron microscopy with polarization analysis, or scanning transmission X-ray microscopy.

Several techniques are suitable to directly probe magnons at the 2D limit (approaches exploiting the coupling of magnons to other excitations are addressed in Sec.~\ref{ssec:excitations}). Inelastic spectroscopies are highly relevant, e.g., Raman scattering spectroscopy, covering the $\sim$100~GHz/100~THz range and capable of detecting ferromagnetic magnons ($\lesssim$1~meV) in single-layer CrI$_3$ \cite{Cenker2021} and antiferromagnetic ones (15-20~meV) in FePS$_3$ tetralayers \cite{Liu2021}, Brillouin light scattering revealing spin waves (60~$\mu$eV) in few-layers Fe$_5$GeTe$_2$ \cite{Sampaio2025}, resonant inelastic X-ray scattering unveiling $\sim$ 300~meV dispersionless spin excitations in single-layer FeS \cite{Pelliciari2021}, or inelastic tunneling spectroscopy detecting magnons (3-7~meV) in bilayer CrI$_3$ \cite{Klein2018}.

Another technique of choice is scanning tunneling microscopy (STM), bringing atomic-scale spatial resolution, high-energy and wave-vector resolution, and potentially too, time-resolution in the THz time-scale. STM inelastic electron tunneling spectroscopy is well established to study magnons on surfaces, and in certain cases allows one to map out the joint density of states in energy/momentum space \cite{Mitra2023,Ganguli2023}, with sensitivity to moir\'{e} patterns that may scatter magnons \cite{Ganguli2023}. 

Finally, it is worth mentioning that the time-varying magnetic fields locally produced by spin waves may be detectable with a diamond's (nitrogen-vacancy) color center \cite{Zhou2021}.

\subsubsection{Magnon transport}

Transport of incoherent magnons in magnetic insulators has emerged as a central concept in spintronics. A workhorse system is the ferrimagnetic insulator yttrium iron garnet (YIG) {contacted to heavy-metal electrodes (e.g., Pt) \cite{Cornelissen2015}}, in which magnons can be generated electrically at one location, diffuse over micrometer distances, and be detected nonlocally at a remote spot, without charge conduction. A charge current excites magnons in two ways: (\textit{i}) electrically, via the spin Hall effect, and (\textit{ii}) thermally, via Joule heating and the spin Seebeck effect. The thus-generated incoherent magnons propagate through the magnet and {are detected after being converted back (via inverse spin Hall effect)} into charge signals. The first- and second-harmonic responses correspond to (\textit{i}) electrically and (\textit{ii}) thermally generated magnons, respectively.

With their low dimensionality and a variety of possible anisotropies, vdW antiferromagnets represent a new platform for exploring magnon transport. The first experimental report, in MnPS$_3$ \cite{Xing2019}, demonstrated thermally driven magnon transport, with diffusion lengths on the micron scale. Later, still with MnPS$_3$, the nonlocal signal could be electrically switched on and off using a gate electrode, effectively realizing a magnon valve, and it was found that magnon transport is sensitive to the spin-flop transition occurring in this material \cite{Feringa2022}. {In the isostructural compound NiPS$_3$, nonlocal second-harmonic signals exhibit discontinuous jumps at the anisotropic spin-flop transition and an inverse-square distance dependence, consistent with a dominant intrinsic spin Seebeck effect \cite{Yuan2025}.}
A second notable system, CrPS$_4$, was used to demonstrate a giant electrically-tunable transport anisotropy \cite{Qi2023} and transport of electrically-excited magnons (something that has remained so far out of reach with other vdW magnets), with relaxation lengths approaching a micron \cite{deWal2023}. Finally, vdW magnets with an easy-plane anisotropy and a weak interlayer exchange interaction are promising candidates for realizing the proposal for a spin superfluid \cite{Takei2014}, where a phase-coherent spin current would propagate without decay. While current experiments probe incoherent, diffusive magnon transport, the panoply of vdW magnets seems well suited to explore the crossover to such coherent spin transport regimes.

It is worth noting that most vdW magnon transport studies to date rely on thermally generated magnons (second-harmonic signals), whereas purely electrical injection (first harmonic) has proven more difficult. This limitation likely stems from poor spin-mixing conductance at the metal/magnet interfaces and shorter diffusion lengths in ultrathin samples, which suppress electrically-driven signals \cite{deWal2023}. Improved interfaces, alternative detectors, or resonant excitation will be essential for advancing from thermally dominated transport to fully controllable, electrically driven magnonic devices in 2D magnets.

\subsubsection{Topological magnons}
\label{ssec:topoMagnons}
 
Magnon topology has been extensively studied theoretically in ferromagnets \cite{Zhang2013f,Shindou2013} and antiferromagnets \cite{Nakata2017}, especially in 2D, with proposals of magnonic analogs to topological insulators and semimetals. {Akin to electronics, non-trivial topology is imprinted in the magnon band structure in the form of gapped Dirac cones. However, because magnons are bosons, thermally excited states across the entire spectrum contribute to magnon, spin and heat transport. Consequently, magnonic systems described as topological “insulators” or “semimetals”---terms referring to the topology of their magnon band structure---are not truly insulating or semimetallic with respect to magnon transport, but they display topologically protected quantized edge states, coexisting with their trivial bulk states.}
  
\begin{figure}[h!]
\includegraphics[width=8cm]{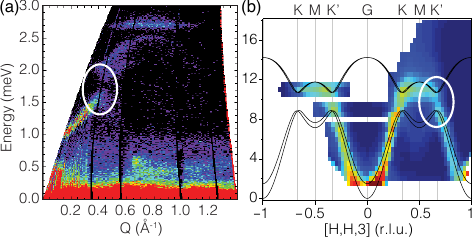}
\caption{Magnon band structure and nontrivial gaps (white circles) measured by inelastic neutron scattering in (a) a kagome ferromagnet (metal-organic framework) and (b) CrSiTe$_3$. From~\cite{Chisnell2015,Zhu2021}. \label{fig:magnongap}}
\end{figure}

Neutron scattering has revealed magnon gaps in a kagome ferromagnet \cite{Chisnell2015}, in CrI$_3$ \cite{Chen2018a} and Cr(Si,Ge)Te$_3$ \cite{Zhu2021} (Fig.~\ref{fig:magnongap}). The gaps were claimed to be topologically nontrivial, although clearly establishing this nontrivial character would require detecting topological edge states. Alternatively, oscillations of the scattering intensity, when circling around the Dirac nodes, as reported in 3D CoTiO$_3$ \cite{Elliot2021,Yuan2020b}, provide key information on the winding of the Dirac isospin, a hallmark of topology. This approach could apply to vdW systems.
 
The other route to demonstrate a nontrivial topology is the observation of quantized edge states. Unlike with electrons \cite{Konig2007,Chang2013}, in the case of magnons these edges states coexist with bulk states and do not lead to a quantized Hall signal. In fact, a magnon Hall effect has been reported in several topologically nontrivial vdW magnets, a kagome ferromagnet \cite{Hirschberger2015b} and CrI$_3$ \cite{Zhang2021c}, but none display any remarkable feature associated with the edge states.  {The only claimed quantization is for the transverse heat conductance of $\alpha$-RuCl$_3$ \cite{Kasahara2018,Yokoi2021}, where the magnetic excitations are expected to be Majorana fermions rather than bosonic magnons.} This observation has not been confirmed independently yet, and is the subject of intense debate \cite{Czajka2023}. Another important feature of the thermal Hall effect, which substantially hinders its pertinence to quantify topological edge modes, is the importance of the phonon Hall effect, as several experiments suggest that, in vdW systems, it is comparable to the magnon Hall effect \cite{Zhang2021c,Xu2023}. Disentangling these contributions is an important challenge for the near future.

Overall, despite compelling band-structure signatures, an unambiguous experimental demonstration of nontrivial magnon topology has remained elusive. Recently, scanning tunneling microscopy measurements on single-layer CrI$_3$ reported magnon edge states, manifested as magnon-assisted tunneling conductance---a potentially transformative development \cite{Zhang2024} that needs to be reproduced. Establishing whether the observed edge modes are truly topologically protected constitutes another key challenge for future experiments.

\subsubsection{Coupling spin excitations with others}
\label{ssec:excitations}

Being highly susceptible to multiple kinds of external control parameters, 2D/vdW magnets represent versatile platforms wherein the energy scales of different types of excitations can be tuned, and thereby resonances between these excitations.

Magneto-elastic effects (see Sec.~\ref{sssec:stress}), {obviously have an impact on or involve phonons}. A first {instance} might be a structural phase transition, concomitant to a magnetic one, whereby the lattice symmetry changes and therefore the nature of phonons too. Additionally, a renormalization of the phonon energy may occur in presence of a magnetic order or at a magnetic phase transition, \cite{Vaclavkova2020}, simply due to the coupling of the spin and (atomic) position degrees of freedom or in relation with spin-orbit effects, while symmetry changes associated to the appearance of magnetic order beyond the atomic period spawn zone-folded phonon modes, readily observed in Raman scattering experiments \cite{McCreary2020,Pawbake2025}. Those effects are not specific to 2D systems, and have been observed years ago in bulk crystals, including vdW ones. They may occur as a function of temperature, pressure, or external magnetic field, as the system goes through various kinds of magnetic phases---certain vdW systems hosting competing magnetic interactions feature particularly rich phase diagrams.

Spin excitations may form hybrid quasi-particles with phonons, so-called magnon-polarons. Ferromagnetic magnons (GHz range) mainly couple to acoustic phonons at nonzero momentum. Such magnon-polarons are usually detected with neutron scattering. Owing to the large single-ion anisotropy of Fe and Co ions, several vdW antiferromagnets instead have magnon gaps above the THz regime, where optical phonons are found. An external magnetic field or pressure allows for tuning the phonon and magnon energies, and brings them in resonance to generate high-energy magnon-polarons \cite{Hay1969,Liu2021,Pawbake2022,Cui2023}. These quasi-particles sometimes reveal subtle aspects, e.g., a chirality-selective magnon-phonon interaction in FePSe$_3$ \cite{Cui2023}. In this compound, the magnon gap matches the phonon energy, so that magnon-polarons appear naturally at low temperature, in the magnetically ordered state. Another related effect is the coupling between a phonon and the two-magnon continuum that can result in Fano resonance for specific phonons, as observed in NiPS$_3$ \cite{Kim2019}, or in an anomalous temperature dependence of particular phonons, in MnPSe$_3$ \cite{Mai2021}.

Magnon-magnon {mode repulsion} has also been investigated in vdW magnets. Such coupling requires the breaking of a symmetry, which is usually done by applying an external magnetic field away from particular magnetic axes (e.g., easy or hard). The characteristic signature is an avoided crossing between the two energy-versus-field magnon branches. This effect has been observed in easy-plane (CrCl$_3$) \cite{MacNeill2019} and tri-axial (CrSBr) \cite{Cham2022} vdW magnets. The coupling strength and the prominence of the avoided crossing can be controlled, for instance by applying uniaxial stress to change the interlayer exchange interaction \cite{Diederich2022}. The {intrinsic} magnon-magnon interaction, besides, can lead to non-linear magnon dynamics, higher magnon harmonics and magnon sum/difference frequency \cite{Diederich2025}. A last consequence of magnon-magnon interactions is the appearance of hydrodynamics physics, a collective behavior with emergent excitation modes such as a low-energy sound modes \cite{RodriguezNieva2022}.

\begin{figure}[h!]
\includegraphics[width=8cm]{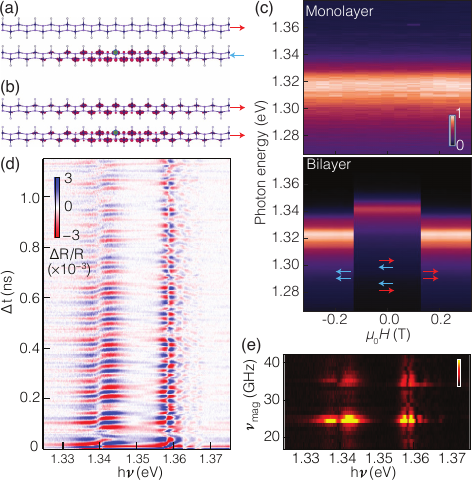}
\caption{Light-magnetism coupling in CrSBr. (a-b) Real-space wave-function, from GW - Bethe Salpeter equations calculations of the lowest-energy exciton in bilayer CrSBr, located around a Cr site, on a side-view for antiferromagnetic (a) and ferromagnetic (b) orders. (c) Exciton line in photoluminescence experiments in the monolayer and bilayer CrSBr, as a function of magnetic field; only the bilayer exciton is influenced by the magnetic order. (d,e) Coupling of excitons to coherent magnons, as observed in transient reflectance experiments; with (e) the Fourier transform of (d). Panels (a-c) have been reproduced from \cite{Wilson2021}, pannels (d,e) from \cite{Bae2022}. \label{fig:CrSBr}}
\end{figure}

A remarkable coupling between electronic band structure and magnetic order has been revealed in a direct-bandgap semiconductor vdW antiferromagnet, CrSBr (Fig.~\ref{fig:CrSBr}). This coupling originates from a spin-dependent electronic interlayer hopping tied to the alternated magnetization of successive layers {[a kind of order also encountered in other semiconducting vdW magnets such as CrI$_3$ or CrPS$_4$ \cite{Zhuang2016,Song2019,Peng2020,Li2020}]}, and it allows for the generation and detection of magnons using ultra-fast optical techniques. This particular hopping produces an effective exciton-magnon interaction manifested as a renormalization of the exciton energy at the magnon frequency \cite{Bae2022,Bae2024}. The dielectric mismatch between the thin CrSBr layer and air, moreover, creates an optical cavity whose eigenmodes couple to the exciton-magnon, a self-hybridization process spawning magnetically tunable exciton polaritons \cite{Dirnberger2023}.

Finally, with the perspective of optical coherent manipulation and sensitive characterization of magnons in minute amounts of matter (ultra-thin vdW material), the coupling between magnons and microwave photons has been demonstrated recently by interfacing a vdW magnet with a superconducting cavity \cite{Zhang2021d,Zollitsch2023}. The coupling is evidenced by an avoided crossing behaviour between the cavity mode and the magnons when they are tuned in resonance by changing an external parameter, and allows for extracting the coupling strength and the Gilbert damping factor.

Most of these interactions have been observed in {relatively thick crystals. This is understandable as in several forefront materials (e.g., CrSBr, CrPS$_4$, CrI$_3$), the rich physics requires several layers for the out-of-plane order to develop, while optical studies of indirect-bandgap vdW semiconductors are least demanding with thick, rather than 2D samples. Direct-bandgap vdW magnets expectedly feature stronger light-matter interaction, and should lend themselves to studies down to the 2D limit.} 

\subsection{Beyond simple collinear spin orders}
\label{ssec:spintextures}

\subsubsection{Non-collinear spin textures, topological or not}

Magnetic orders beyond simple ferro-, ferri-, and antiferromagnetism, whether commensurate or not, collinear or not, arise from \textit{competing} interactions, notably symmetric (e.g., Heisenberg, four-spin) and antisymmetric exchange, such as DMI (Sec.~\ref{ssec:micro}). Spin spirals and helixes, for instance, have been observed and studied for decades \cite{Yoshimori1959,Herpin1959,Dzyaloshinskii1964,Dzyaloshinskii1965b} [for a modern, in-depth discussion, see \cite{Batista2016}]. The case of 2D spin lattices has been extensively documented too, with frequent reference to vdW materials early on \cite{Rastelli1979}. More recently, spin spirals stabilized by DMI were reported in systems with broken inversion symmetry, such as Mn/W(110) interfaces or bulk Fe$_{0.5}$Co$_{0.5}$Si \cite{Bode2007,Yu2010,Heinze2011}. Noncollinear textures also appear as localized magnetic defects---chiral and achiral domain walls, vortices, or bubbles--- within an otherwise homogeneous background. Remarkably, some of these metastable defects (skyrmions, hopfions, etc) exhibit topological robustness. These solitons are classified using homotopy groups \cite{Mermin1979,Braun2012} and are each associated with a topological invariant (or charge), which counts the winding number of the 3D spin field projected onto a plane. Magnetic skyrmions and their siblings (hopfions, bimerons, etc) have been reported in metallic multilayers \cite{Chen2015b,Jiang2015} and noncentrosymmetric magnets \cite{Muhlbauer2009,Yu2010} and are objects of considerable interest in spintronics \cite{Zang2018,Tokura2020}.

Noncollinear spin textures are not unheard of in vdW systems, and canted-spin-phases have been discussed decades ago in NiBr$_2$ \cite{Morimoto1970,Day1976} while spiral spin textures are well-known in FeOCl \cite{Hwang2000,Schaller2022}. More recently, a proximate 3Q antiferromagnetic state has been reported in bulk Co$_{1/2}$TaS$_2$ \cite{Park2022,Park2025b}. There is currently renewed interest in exploring how these textures evolve when thinning down the vdW crystals \cite{Bikaljevic2021,Bao2022}.

{Thin films of vdW magnets have been shown to host skyrmion(ic) phases. It often remains unclear, however,} whether the observed topological objects are genuine skyrmions (driven by DMI–exchange competition) or magnetic bubbles (set by dipolar–exchange balance), what their precise spin texture is, and what fundamentally new features vdW systems bring. First reports revealed bubbles in Fe$_3$GeTe$_2$ \cite{Ding2020}, implemented the virtual-magnetic-field trick to circumvent the need for an external field \cite{Yang2020b}, and pointed to an interface-related DMI in Fe$_3$GeTe$_2$ placed onto WTe$_2$ (a material with high spin-orbit interaction), hence to true skyrmions \cite{Wu2020}. The latter case illustrates the latitude vdW systems offer in terms of stacking and interface engineering  (Fig.~\ref{fig:timeline}). Another possible origin for skyrmion phases was envisaged in intrinsically noncentrosymmetric vdW crystals (or partly vdW): vacancies in the crystal \cite{Chakraborty2022}, or intercalated magnetic ions \cite{Saha2022}, were argued to produce a nonzero DMI whose magnitude and effect can be tuned by, e.g., the amount of intercalants \cite{Liu2025}. More complex skyrmionic textures were also scrutinized \cite{Powalla2023,Zhang2024}. At least to some extent, {and this is remarkable}, vdW materials seem to rapidly catch up with more traditional ones, in terms of temperature stability of skyrmion phases \cite{Liu2024}. {The role of the reduced dimensionality, however, has hardly been touched upon experimentally; more complex 3D topological objects, currently investigated in other classes of materials \cite{Rybakov2022}, have not been reported in vdW magnets (there, reducing their thickness does not seem relevant).}

The DMI might be tunable by an electric field, as the latter disymmetrizes the crystal structure (atomic bonds hold a certain ionic character) \cite{Liu2018b}, which, in the case of 2D materials, may be achieved using electrostatic gates or interfacing a 2D magnet with a 2D ferroelectric \cite{Sun2020}. {Alternatively, inversion symmetry may be intrinsically broken in Janus 2D magnets} (see Sec.~\ref{ssec:compo}). Mechanical deformations offer yet another way to control DMI \cite{Li2022,Zhou2025,Jin2025}---just like other spin interactions.

In the absence of DMI, topological defects usually come in opposite chiralities. Pairs of defects with opposite chirality accompany certain phase transitions \cite{Kasteleyn1963,Kosterlitz1973,Bak1982}, and which kinds of such pairs [e.g., vortex/antivortex, meron/antimeron \cite{Lu2020}] can actually exist and be detected or even directly observed in 2D magnets, are open questions.

\subsubsection{Frustrated systems}

{The notion of frustration was introduced by Toulouse (inspired by Anderson), in Ising spin glasses \cite{Toulouse1977,Vannimenus1977}, where competing interactions prevent magnetic order and are conducive of extensive ground-state degeneracy. Toulouse's concept was later generalized to XY and Heisenberg magnets via a constraint function, defined as the ratio of the ground-state energy to that obtained by independently minimizing all pairwise interactions.} \cite{Lacorre1987}. {The degree of frustration is also measured by the system's residual (nonzero) entropy down to low temperature, following Pauling's arguments on water ice} \cite{Pauling1935}. {Yet another estimator of frustration is the ratio $f$ between the ordering (or freezing) and Curie-Weiss temperatures \cite{Ramirez1994}, with a large ratio indicating strong suppression of magnetic order.}

{Today, highly frustrated magnetism \cite{Lacroix2011} remains a rapidly developing field. It stems} from competing interactions (as in spin glasses), or geometrical frustration, in structurally ordered magnets \cite{Moessner2006}. The distinction is however not strict, and the term `frustrated' has sometimes been applied too broadly, including to systems whose unusual features derive from those of a parent, truly frustrated model. Note that, competition between interactions is \textit{not} sufficient to produce frustration{, and generally only yields conventional kinds of orders}. True (geometrical) frustration arises only for specific combinations of lattice geometry, spin symmetry (Ising, XY, Heisenberg), and interaction type (ferro-/antiferromagnetic).

{In 2D, even the simplest model, the archetypical short-range triangular Ising antiferromagnet (Sec.~\ref{ssec:stat}) that exhibits no long-range order at any temperature, eludes frustration in practice, including in vdW MX$_2$ compounds, due to small lattice distortions, interlayer coupling or non-purely Ising spins} \cite{Collins1997}. The \textit{Heisenberg} triangular antiferromagnet has a {predicted} continuously degenerate ground state with a $\mathbb{Z}_2$ chiral symmetry \cite{Kawamura1984}{, yet to be realized experimentally}. Strikingly, here frustration does not prevent long-range order—unlike the unfrustrated 2D Heisenberg ferromagnet, which cannot order according to Mermin–Wagner theorem.
  
{Can 2D/vdW magnets, particularly the recently discovered and truly 2D ones, represent new candidates for frustrated magnetism?} Large $f$ ratios have been found in several vdW magnets, M$_2$T$_2$X$_5$ (M, a transition metal, T a triel atom, X a chalcogen) \cite{Nakatsuji2007,Shen2022,Williams2024}, 2D VSe$_2$ \cite{Wong2019} and metal halide alloys \cite{Tartaglia2020}. A large $f$, unfortunately, is not unique to frustrated systems, and relating it to a residual entropy density would be a better diagnostic. Complementarily, a disordered, liquid-like manifold should appear in neutron scattering as the absence of magnetic Bragg peaks at low temperature \cite{Mirebeau2014}, although the technique remains unsuitable for single layers. 

Is there a specific interest or advantage in studying vdW frustrated magnets? We believe the answer is yes, despite the impressive amount of existing results in highly frustrated magnets obtained for about three decades. First, vdW materials provide a platform to explore both spin-liquid and spin-glass physics, with the unique possibility of tuning structural disorder—e.g., via intercalation, which is specific to lamellar systems{, as recently demonstrated \cite{Pal2026}}. Defect-free vdW crystals also broaden the landscape, as they can be metallic, insulating, or semiconducting, enabling coupling between spin and other degrees of freedom. Approaching the monolayer limit may reveal new frustration mechanisms, such as the `orphan' bonds predicted in 2D spin-ice slabs \cite{Jaubert2017}---experiments here remain difficult in conventional frustrated magnets. Truly 2D platforms would also open unexplored territory for classical and quantum frustrated models, potentially advancing the search for quantum spin liquids \cite{Knolle2019}. Finally, moir\'{e} engineering may enable spatial modulation of exchange interactions, offering access to novel (frustrated) spin models.

\subsection{Orbital magnetism}
\label{ssec:orbital}

Orbital magnetism is central to understanding and designing materials with large perpendicular magnetic anisotropy [see, e.g., \cite{Weller1995}]. It is a key ingredient of the anomalous Hall effect \cite{Nagaosa2010}, and is at the heart of so-called `orbitronics' phenomena, wherein orbital currents influence the magnetic order of magnetic thin films \cite{Go2021}. The strength of the orbital magnetic moment is set by the interplay between the chemical environment, the dimensionality, and the spin-orbit interaction. As a reference, in bulk Co it amounts to $\simeq$0.15 $\mu_B$, i.e., about 10\% of the spin moment \cite{Chen1995}. This ratio may increase up to 20\% due to interface effects \cite{Nistor2011}. In cubic transition metal compounds, the orbital moment is quenched due to the orbital degeneracy of the $t_{2g}$ states, while again, the reduction of dimensionality at an interface may switch it on \cite{Weller1995}. In 2D magnets, the confinement of the electron motion may favor a large orbital moment normal to the plane.

\begin{figure}[h!]
\includegraphics[width=8cm]{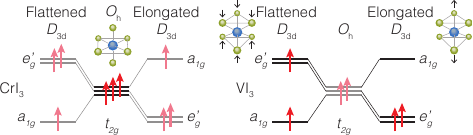}
\caption{Electronic levels of CrI$_3$ and VI$_3$. In CrI$_3$, there is no distortion; the three electrons occupy the $t_{2g}$ state, quenching the orbital moment ($L$). In VI$_3$, trigonal distortion results in the coexistence of two configurations, one $L$-polarized and the other $L$-quenched, resulting in an overall large $L$. \label{fig:vi3}}
\end{figure}

However, in the majority of 2D magnets considered so far, the orbital moment was found to be small, 0.1~$\mu_B$ or less \cite{Ovesen2024}. It is about -0.03~$\mu_B$ in CrSiTe$_3$ \cite{Fujita2022}, -0.07~$\mu_B$ in CrI$_3$ \cite{Frisk2018} and 0.07~$\mu_B$ in CrPS$_4$ \cite{Buccoliero2026}. In contrast, a giant orbital moment of 0.6 $\mu_B$ has been reported in VI$_3$ \cite{Hovancik2023,DeVita2024}. Why two iso-structural compounds, CrI$_3$ and VI$_3$, have so different orbital moments may be appreciated within a simplified picture. Their magnetic ions (Cr, V) both experience a proximate octahedral crystal field ($O_h$). The electron count in the $t_{2g}$ state, however, differs. It is three in CrI$_3$, leading to $L=0$ in the lowest approximation. In contrast, in VI$_3$, a trigonal distortion splits the $t_{2g}$ level into a singly-degenerate $a_{1g}$ and doubly-degenerate $e'_g$ levels (Fig.~\ref{fig:vi3}). The two electrons can assume two possible configurations. In case of trigonal elongation, both electrons occupy the doubly-degenerate $e'_g$ level, resulting in orbital quenching. In case of trigonal flattening, only one electron occupies the singly-degenerate $a_{1g}$ state, hence $L=1$ \cite{Sandratskii2021,Nguyen2021}. The coexistence of these two configurations leads to a reduced, though quite large, orbital moment \cite{Hovancik2023}. The microscopic origin is distinctive from that in CoO for instance, where the spin-orbit coupling alters the orbitally-degenerate $t_{1g}$ state \cite{Jauch2002}. The direct consequence of VI$_3$'s giant orbital moment is the observation of Kitaev interactions \cite{Gu2024}.

A large orbital moment has also been reported in truly 2D systems, like FePS$_3$ \cite{Lee2023} and NiBr$_2$ \cite{Bikaljevic2021} antiferromagnets, as well as bulk Fe$_{1/4}$TaS$_2$, displaying an orbital moment of 1~$\mu_B$ \cite{Ko2011}. A high-throughput investigation across candidate 2D magnets suggests that materials composed of 5$d$ elements tend to exhibit the largest orbital moment, up to 0.5~$\mu_B$ for PtCl$_2$\comment{, without correlation to the magnetic anisotropy in general} \cite{Ovesen2024}\comment{, which is not surprising as there is no direct relationship between both effects}.

Estimates of the orbital moment and its microscopic origin rely on spectroscopic data interpreted with multiplet theory and \textit{ab initio} calculations. While a sizable orbital contribution is consistently inferred in CoO, VI$_3$, FePS$_3$, etc., its quantitative value and microscopic interpretation remain uncertain due to methodological approximations. Improved spectroscopic resolution, notably via RIXS, and more accurate multiplet treatments are required for a reliable description. An interesting route to explore is the search for stacking and interfaces that could promote orbital magnetism either by the proximity effect or by stabilizing the appropriate structural phase. As illustrated in the case of VI$_3$, the energy difference between the two coexisting phases can be very small (5~meV), so favoring one configuration over the other could lead to the realization of a giant orbital moment. Such strategies would be of great importance for the realization of orbital torque \cite{Lee2021b,Gupta2025} and orbital pumping effects \cite{Hayashi2024,Keller2025}.

\subsection{Van der Waals junctions for spin transport}
\label{ssec:transport}

{From the inception of graphene research, magneto-transport \cite{Novoselov2004,Novoselov2005b,Zhang2005} and magneto-resistance measurements \cite{Hill2006,Tombros2007} proved invaluable to explore the (spin-dependent) properties of truly 2D materials, often exploiting their optimal amenability to the electric-field effect. Device architectures have diversified, adapting designs} from conventional metallic spintronics, e.g., magnetic tunnel junctions, spin filters, and spin valves. We deliberately omit spin–charge interconversion studies in vdW monolayers and heterostructures interfaced with traditional magnets, refering the reader to dedicated reviews \cite{Galceran2021,Sierra2021}.
 
Vertical spin filters were among the first spintronic devices explored with 2D/vdW magnets. Conventional magnetic tunnel junctions using vdW magnetic electrodes (e.g., Fe$_3$GeTe$_2$) separated by a few-nanometer $h$-BN barrier were first investigated. These devices exhibit impressive magnetoresistance---up to several hundred percent at cryogenic temperature---thanks to excellent interfacial crystallinity and minimal amount of defects \cite{Wang2018d}, yet room-temperature values are still below those iof standard CoFeB/MgO/CoFeB junctions \cite{Ikeda2008}. The magnetoresistance is tunable with applied bias \cite{Min2022}, a feature that has long been observed in conventional magnetic tunnel junctions \cite{Teresa1999,Bowen2005}. With oxide tunnel barriers however, the accessible bias range is typically limited to 1-2~V, beyond which magnetoresistance strongly degrades or dielectric breakdown occurs. Tunnel junctions with vdW materials sustain comparatively much larger bias voltages [-4~V to +4~V \cite{Min2022}], giving access to richer bias dependence of the tunneling magnetoresistance that reflects the electronic band structure of the system.

{A rather unconventional kind of vertical architecture has been introduced with} \textit{bilayer} CrI$_3$. {Here there is no physical insulating barrier like in standard spin filters, but only the vdW gap between two magnetic layers; still their magnetization can be switched independently. Such behavior is also found in other vdW antiferromagnets \cite{Telford2020}, but not systematically, depending on the kind of transport involved (space-charge- vs injection-limited) \cite{Shi2024,Lin2025}}.  At low fields, antiferromagnetic coupling between the layers creates a large tunnel barrier and low conductance, while higher fields align the spins, reduce the barrier, and increase conductivity. Details depend on the number of magnetic layers and their internal coupling{, which itself depends on the twist angle between layers \cite{BoixConstant2024,Chen2024}}. Below the (low) ordering temperature, the conductivity changes are enormous, reaching almost 20,000\% \cite{Song2018,Klein2018,Wang2018c}, beyond what is obtained with more traditional systems \cite{Miao2009}. Besides record performances, we believe spin filters provide a direct way to probe magnetic configurations in 2D/vdW magnetic insulators and enable devices such as spin-tunnel field-effect transistors, which would be extremely challenging to realize without the stacking flexibility of vdW or 2D/3D interfaces \cite{Jiang2019}.

Lateral spin valves, combining spin injection, transport and collection, have been studied intensively over the past 30 years, aiming at a spin field-effect transistor \cite{Datta1990}---though with modest success \cite{Koo2009,Choi2018}---or, since the advent of graphene, long-range interconnects \cite{Tombros2007,Han2011,Yang2011} or switches \cite{Yan2016}. Such lateral spin valves have also extended to 2D/vdW magnets using, e.g., metallic Fe$_5$GeTe$_2$ or Fe$_3$GaTe$_2$ as spin sources for injection into graphene channels \cite{Pan2023,Zhao2023b}. Further exploiting the special properties of 2D/vdW magnets and their heterostructures, insulating CrGeTe$_3$ was used to implement magnetic proximity with graphene (Fig.~\ref{fig:timeline}). Predetermined sections of a graphene nonlocal spin valve were {spin-polarized by magnetic proximity}, acting as a spin injector and detector on a seamless 2D/vdW device \cite{Yang2025}, preserving graphene's advantageous spin transport properties.

{Spin-orbit torque (SOT) devices rely on the manipulation of magnetization within a layer by the spin current through a neighboring layer, via a strong spin-orbit coupling. Such memory devices have been realized with 2D/vdW magnets, first some} having perpendicular magnetic anisotropy, e.g., Fe$_3$GeTe$_2$, in contact with a metallic layer such as Pt \cite{Wang2019,Alghamdi2019}. {Further sophistication exploited canted-magnetization 2D magnets, to achieve} field-free SOT magnetization switching, {remarkably at room temperature} \cite{Zhao2025}. {Even then, the advantage with respect to mature technologies exploiting more traditional metallic thin films, is (so far) not obvious}. 

{Besides the materials variety offered by 2D stacking \cite{Kao2022,Shin2022}, unique SOT phenomena are at reach with 2D/vdW magnets. (\textit{i}) Self-torque, arising from the low crystal symmetry and reported in Fe$_3$GeTe$_2$, enables intrinsic current-driven magnetization switching, without an additional strong spin-orbit layer\cite{Zhang2021b,Johansen2019,Smaili2021}. (\textit{ii}) Interlayer coupling in vdW stacks or moir\'{e}s may lead to novel spin dynamics. (\textit{iii}) The electric-field-effect susceptibility can be used to boost or quench SOT, e.g., in a three-terminal device \cite{Vojacek2024}.}

\subsection{Topology effects and quantum geometry}
\label{ssec:topoQgeom}

The realization of topological phases and quantum geometry in truly 2D materials remains in its infancy, despite enormous theoretical progress. It seems that most topological effects require material purity that are still out of reach in 2D/vdW magnets, and only found in graphene wherein emergent quantum topology, such as a 2D $\mathbb{Z}$ topological insulator, has been reported (see discussion in Sec.~\ref{ssec:beyond}). Over the past year, strong candidates have been identified. In Sec.~\ref{ssec:topoMagnons}, we mentioned the recent report of topological edge magnons in monolayer CrI$_3$ \cite{Zhang2024b}. The few-septuple-layer MnBi$_2$Te$_4$ antiferromagnet is also attracting considerable interest \cite{Otrokov2019a}. It exhibits a quantized magnetoelectric coupling \cite{Liu2020} and hosts an anomalous nonlinear transport effect linked to the quantum metric \cite{Gao2023,Wang2023b}, a phenomenon also observed in CrSBr \cite{Jo2025}. Recently, oscillations of the magnetoelectric coupling driven by coherent antiferromagnetic magnons---interpreted as a signature of axion-like electrodynamics---have been reported in six-septuple-layer MnBi$_2$Te$_4$ \cite{Qiu2025}. These exciting results call for further investigations, and their connection to quantum geometry and topology must be challenged. However, considering the remarkable recent progress in materials synthesis and interfacial engineering, {observing such exotic phenomena in truly 2D magnets or carefully designed heterostructures shall be (soon) at reach.}

\subsection{Beyond 2D/vdW magnets: flatband magnetism}
\label{ssec:beyond}

We so far left aside two important classes of 2D magnetic systems whose magnetism is not associated with magnetic elements, but rather with electron correlations in flat bands arising from the destructive interference of delocalized Bloch states \cite{Leykam2018,Aoki2025}. Due to their low velocity and high density of states, flat bands with degeneracies---for example, in spin, valley, or layer degrees of freedom---are particularly susceptible to electron-electron interactions, which can lift these degeneracies and induce spontaneous symmetry breaking associated with the corresponding degrees of freedom. This leads, for example, to quantum Hall ferromagnetism \cite{Nomura2006}, as early observed in semiconductor quantum wells \cite{Barrett1995,Schmeller1995}, and, more recently, in mono- and multilayer graphene \cite{Feldman2009,Zhao2010,Young2012}. 

{More complex orders can be obtained in graphene's $N=0$ Landau level, by adjusting the charge carrier density, e.g., canted antiferromagnetism} \cite{Young2012}. Remarkably, the connection with spintronics and condensed matter magnetism has strengthened over the years as quantum Hall skyrmions \cite{Zhou2020,Atteia2021}, magnon transport \cite{Wei2018,Assouline2021}, and magnon superfluidity have been unveiled \cite{Stepanov2018,Takei2016}. The implementation of spintronics detection schemes (nonlocal spin valves, tunnel junctions, and spin-torque devices) {provides new opportunities to probe and harvest} such exotic excitations. Yet, realizing these phases of matter still requires extreme conditions—ultraclean samples, magnetic fields of several Tesla, and sub-Kelvin temperatures—, currently limiting their exploration.

Other magnetic systems covered by the present paper are moir\'{e} systems obtained by twisting adjacent layers (which are, individually, nonmagnetic). There also, flat bands develop due to quantum interference, as mini-bands in the band structure. The thus-emerging correlated phases \cite{Bistritzer2011}, among which magnetic ones \cite{Sharpe2019,Serlin2020}, have been covered already in a number of reviews [e.g., \cite{Mak2022}]. Here we only note that the techniques developed in the context of 2D magnetism and spin transport, from surface engineering to spin-valve and spin torque physics and technology, will be highly relevant to explore and manipulate these new magnetic phases. {Finally, several lattice geometries (e.g., kagome, Lieb, dice) intrinsically host flat bands, and we anticipate that implementing them in 2D materials shall be a route to 2D quantum magnetism.}

\section{Final remarks}
\label{sec:conclusion}
{The isolation of magnetic 2D crystals in 2016 and 2017 raised great hopes for gaining new insights into longstanding and emerging questions alike, extending far beyond what other quasi-2D systems---surface-supported metallic layers and lamellar bulk crystals---had previously made possible. Have these expectations been met? What truly disruptive perspectives do these systems open? }

\noindent {$\bullet$ To date, obtaining \textit{clearcut signature of 2D magnetism}, in the statistical physics sense, remains exceptionally rare; famous theorems and scenarios (Kosterlitz-Thouless, Mermin-Wagner, etc) are still hard to test experimentally. A plethora of phenomena wherein two-dimensionality is expectedly conducive of unconventional properties, tied to topological or strong correlations or orbital effects, have hardly been scrutinized in 2D magnets.}

\noindent {$\bullet$ Much of today's `2D magnetism' research still focuses on \textit{bulk vdW crystals} rather than genuinely 2D systems. Misuse of the term `2D' has become widespread, even for materials that retain essentially 3D behavior. True 2D magnets remain difficult to study because of their instability and the extreme sensitivity required for their characterization. Nevertheless, renewed interest in thicker (i.e., non-2D) vdW magnets---many known for decades---has been facilitated by key progresses in magnetic-sensitive techniques and micro/nano-fabrication, and still proved highly rewarding.}

\noindent {$\bullet$ Among quasi-2D candidates, \textit{layered semiconductor antiferromagnets} have emerged as especially amenable to optically-controllable spin excitations and interlayer magnetic phenomena. Few-layer and heterostructured magnets exhibit striking sensitivity to \textit{electric fields, stacking order, twisting, and moir\'{e} engineering}.}

\noindent {$\bullet$ The evolution of the field towards \textit{advanced functionalities} is rather natural. Pressure effects on magnons and magnetic orders can be strong and anisotropic, although not unique to 2D/vdW magnets. Whether stronger magnetic/light-emission proximity effects can be achieved is yet to be fully figured out, and the potential of these structures for achieving topological magnons remains to be proven. Entire classes of phenomena---including magnetic frustration, quantum geometry, curved magnetism, and topology-driven effects---remain \textit{largely unexplored in 2D magnets}.}

\noindent {$\bullet$ Several conventional effects have been transposed to 2D/vdW magnets. Their symmetries, atomic structure, and stacking versatility are promising for stabilizing non-collinear, sometimes topological, spin textures, or to manipulate magnetization; spin-dependent transport (magnetoresistance) occasionally shows very pronounced effects, some of which in unprecedentedly-thin stacks, sometimes fully controlled electrically, however often at \textit{lower operation temperature than with technologies relying on more mature materials}.}

\noindent {$\bullet$ The \textit{field has advanced rapidly}, but often at the expense of rigor and reproducibility. Numerous high-profile claims in the literature have proven exaggerated, weakly supported, or outright incorrect. To name a few misleading elements, beyond the widespread misuse of the term `2D', let us mention above-room-temperature intrinsic ferromagnetism in several compounds, room-temperature magnetism achieved in certain compounds by electric-field effect, assumed shape effects in stabilizing ferromagnetism above room-temperature and overlooked irradiation damage, room-temperature (again!) diluted ferromagnetism, magnetic properties assigned to materials that actually do not have the believed structure/composition, unrealistically-strong interlayer spin-spin interactions in modeling---not to mention ill-supported interpretations, or unconvincing analyses of, e.g., scaling laws. We warn that several review papers simply ignore these flaws, and even make multiple untrustworthy/unfaithful references to previous works.}


Let us finally not forget that the field is still relatively new, and is meant to progress and hopefully grow more mature in the coming years. The field remains scientifically exciting, but it urgently requires greater experimental rigor, theoretical maturity, and critical scrutiny.

\section{Acknowledgments}

We thank the Referees for their insightful advices, which helped us delineate some of the most distinctive aspects of modern studies on van der Waals and 2D magnets, while contextualizing these studies in the perspective of older lines of works. J.C., C.R., N.R., A.M., and C.F. acknowledge financial support from France 2030 government investment plan managed by the French National Research Agency under grant reference PEPR SPIN --- SPINMAT ANR-22-EXSP-0007, SPINCHARAC ANR-22-EXSP-0008, SPINTHEORY ANR-22-EXSP-0009, MULTISPIN ANR-24-EXSP-0008, as well as ANR-23-CE09-0034-03 “NEXT”. C.F. acknowledges support from projects ANR-23-QUAC-0004 and CEFIPRA 7104-2. C.R. acknowledges support from ESR/EquipEx+ 2DMAG (grant ANR-21-ESRE-0025) and fruitful discussions with Thomas Guillet. L.E.H. and F.C. acknowledge funding from MICIU/AEI/10.13039/501100011033 (Grant No. CEX2020-001038-M) and from MICIU/AEI and ERDF/EU (Project No. PID2024-155708OB-I00).

\bibliography{2D_vdW_mag.bib}

\end{document}